\documentclass{aa}  

\usepackage{graphicx}
\usepackage{amssymb,amsmath}
\usepackage{latexsym}
\usepackage{mathtools}
\usepackage{epsfig}
\usepackage{natbib}
\usepackage{txfonts}
\begin{document}

   \title{Long-term evolution of the Galilean satellites: the capture of Callisto into resonance}
   \titlerunning{Long-term evolution of the Galilean satellites}
   \author{Giacomo Lari\inst{1}
          \and
          Melaine Saillenfest\inst{2}
          \and
          Marco Fenucci\inst{1,3}
          }
   \authorrunning{G. Lari et al.}
   \institute{Department of Mathematics, University of Pisa,
              Largo Bruno Pontecorvo 5, 56127 Pisa, Italy\\
              \email{lari@mail.dm.unipi.it}
         \and
             IMCCE, Observatoire de Paris, PSL Research University, CNRS, Sorbonne Universit\'e, Universit\'e de Lille, 75014 Paris, France
         \and
             Department of Astronomy, Faculty of Mathematics, University of Belgrade, Studentski trg 16, 11000 Belgrade, Serbia
             }
   \date{Received --- / Accepted ---}


  \abstract
  {The Galilean satellites have very complex orbital dynamics due to the mean-motion resonances and the tidal forces acting in the system. The strong dissipation in the couple Jupiter--Io is spread to all the moons involved in the so-called Laplace resonance (Io, Europa, and Ganymede), leading to a migration of their orbits.}
  {We aim to characterize the future behavior of the Galilean satellites over the Solar System lifetime and to quantify the stability of the Laplace resonance. Tidal dissipation permits the satellites to exit from the current resonances or be captured into new ones, causing large variation in the moons' orbital elements. In particular, we want to investigate the possible capture of Callisto into resonance.}
  {We performed hundreds of propagations using an improved version of a recent semi-analytical model. As Ganymede moves outwards, it approaches the 2:1 resonance with Callisto, inducing a temporary chaotic motion in the system. For this reason, we draw a statistical picture of the outcome of the resonant encounter.}
  {The system can settle into two distinct outcomes: (A) a chain of three 2:1 two-body resonances (Io--Europa, Europa--Ganymede, and Ganymede--Callisto), or (B) a resonant chain involving the 2:1 two-body resonance Io--Europa plus at least one pure 4:2:1 three-body resonance, most frequently between Europa, Ganymede, and Callisto. In case A  (56\% of the simulations),  the Laplace resonance is always preserved and the eccentricities remain confined to small values below 0.01. In case B (44\% of the simulations), the Laplace resonance is generally disrupted and the eccentricities of Ganymede and Callisto can increase up to about 0.1, making this configuration unstable and driving the system into new resonances. In all cases, Callisto starts to migrate outward, pushed by the resonant action of the other moons.}
  {From our results, the capture of Callisto into resonance appears to be extremely likely (100\% of our simulations). The exact timing of its entrance into resonance depends on the precise rate of energy dissipation in the system. Assuming the most recent estimate of the dissipation between Io and Jupiter, the resonant encounter happens at about 1.5 Gyr from now. Therefore, the stability of the Laplace resonance as we know it today is guaranteed at least up to about 1.5 Gyr.}



   \maketitle


\section{Introduction}\label{sec:intro}


The Galilean satellites are the four biggest moons of Jupiter, discovered by Galileo Galilei in 1610. Ordered with respect to their distance from Jupiter, they are: Io (1), Europa (2), Ganymede (3), and Callisto (4). In 1798, Laplace observed that the mean motions of Io, Europa, and Ganymede are in 4:2:1 commensurability. This configuration is made of two 2:1 two-body mean-motion resonances involving the couples Io--Europa and Europa--Ganymede. Using $\lambda_i$ to denote the mean longitude of the $i$th satellite and $\varpi_i$ its longitude of pericenter, we currently have
\begin{equation}\label{lap2bod}
   \begin{aligned}
      \lambda_1-2\lambda_2+\varpi_1&\sim 0 \,,\\
      \lambda_1-2\lambda_2+\varpi_2&\sim \pi \,,\\
      \lambda_2-2\lambda_3+\varpi_2&\sim 0 \,,
   \end{aligned}
\end{equation}
where $\sim$ stands for ``closely oscillates around''. Combining the last two relations, we obtain:
\begin{equation}
   \lambda_1-3\lambda_2+2\lambda_3\sim \pi \,,
   \label{lapang}
\end{equation}
which involves the mean longitudes of all three satellites. This relation is commonly known as the ``Laplace resonance''. Moreover, from the first two relations in Eq.~\eqref{lap2bod}, we can note that
\begin{equation}\label{antialign}
   \varpi_1-\varpi_2\sim \pi \,,
\end{equation}
which implies that the orbits of Io and Europa are anti-aligned.

The orbits of regular satellites in the Solar System are generally the result of billions of years of dynamical evolution. Tidal forces between the satellites and their host planet produce dissipative effects that lead to a radial migration of the satellites over long timescales. Tidal dissipation in the Galilean satellites is the source of spectacular phenomena like the volcanism on the surface of Io \citep{PEALE-etal_1979}, or the preservation of oceans of liquid water under the icy crust of Europa \citep{CASSEN-etal_1979} and probably Ganymede.


The formation of resonant configurations between satellites remained a mystery for a long time, until \cite{GOLDREICH_1965} put forward the idea of resonance capture through dissipative migration. Numerous works further studied this mechanism and its application to the satellites of Jupiter and Saturn, confirming the extreme importance of tidal dissipation in their long-term evolution (see e.g., \citealp{SINCLAIR_1972,GREENBERG_1973,SINCLAIR_1975}). More detailed scenarios were then developed for the Galilean satellites. \cite{YODER_1979} and \cite{YODER-PEALE_1981} suggested that the migration of Io has always been faster than the migration of the other Galilean satellites; as a result, Io first captured Europa into mean-motion resonance, which sped up its migration and led to the subsequent capture of Ganymede. These latter authors estimated the respective probabilities of each resonance capture, and deduced lower and upper bounds for the tidal dissipation within Jupiter. \cite{TITTEMORE_1990} showed that the establishment of the Laplace resonance has probably been preceded by a chaotic phase in which the eccentricities of Europa and Ganymede increased dramatically. This induced a high tidal friction within the two bodies, which could explain why Ganymede and Callisto have very different surface properties. The same scenario was proposed by \cite{MALHOTRA_1991} and \cite{SHOWMAN-MALHOTRA_1997}, who showed that the chaotic phase was most likely due to the crossing of three-body mean-motion resonances between Io, Europa, and Ganymede (contrary to what had first been announced by \citealp{TITTEMORE_1990}). They used this argument to obtain refined bounds for the dissipation parameters.

Other works suggest that the Laplace resonance settled during the formation of the Galilean satellites. \cite{GREENBERG_1982} conjectured that the satellites were originally in deep resonance and that they are currently evolving out of resonance. In this scenario, the forced eccentricities of the satellites were initially much higher, with a consequent stronger tidal friction within all three resonant moons. \cite{GREENBERG_1987} found a path of stable configurations leading from the current configuration of the Laplace resonance back to deeper resonant states. Forced to follow this path because of tidal dissipation, the Laplace angle would then have passed through asymmetric equilibria (i.e., different from $0$ or $\pi$). Nevertheless, this scenario gives no information about how primordial the Laplace resonance is. According to \cite{PEALE-LEE_2002}, the Laplace resonance could have settled during the formation of the satellites in the circumjovian disk as a result of differential migration. Following the scenario depicted by \cite{CANUP-WARD_2002}, Ganymede underwent the faster Type-I migration because of its larger mass. Moving toward Jupiter, it first captured Europa in resonance, followed by Io, and this process happened relatively quickly (about $10^5$ years). After the dissipation of the disk, the satellites then reached their current state by tidal dissipation.


The future evolution of the Galilean satellites has received little attention so far. Over short and medium timescales (up to $10^5$ years), the stability of the Laplace resonance has been confirmed by \cite{MUSOTTO-etal_2002} and \cite{CELLETTI-etal_2019}, however little is known about its stability as a result of tidal dissipation over long timescales. It is not clear whether new reorganizations of the orbits of the moons are to be expected, as found for instance in exoplanetary systems \citep{BATYGIN-MORBIDELLI_2013,PICHIERRI-etal_2019}. Because of the mean-motion resonances in Eqs.~\eqref{lap2bod} and \eqref{lapang}, the strong dissipative effects acting on Io \citep{LAINEY-etal_2009} are redistributed among Europa and Ganymede. This implies that the satellites still migrate today, and that important events like the ones that occurred in the past may happen in the future. In particular, Callisto is not currently involved in any mean-motion resonance, which leads us to the question of whether or not the dissipation could ever make it cross a resonance with another Galilean satellite. Since the tidal dissipation produces an outward migration of Io, Europa, and Ganymede \citep{FULLER-etal_2016}, the first important resonance that could be encountered is the 2:1 commensurability between Callisto and Ganymede. From numerous studies of other moons (e.g., \citealp{TITTEMORE-WISDOM_1990,MEYER-WISDOM_2008}) and exoplanets (e.g., \citealp{BATYGIN_2015,CHARALAMBOUS-etal_2018}), we know that a large variety of outcomes are possible, even including the ejection of one satellite (e.g., \citealp{POLYCARPE-etal_2018}).


In this article, we aim to measure the stability of the Laplace resonance over a billion-year timescale under the effects of tidal dissipation. We also aim to determine the possible outcomes of the resonant encounter with Callisto and to quantify its capture probability.


Starting with the works of Lagrange and Laplace, the first comprehensive theories of the orbital dynamics of the Galilean satellites were meant to reproduce their current motion with a high accuracy. These theories were first analytical (e.g., \citealp{SOUILLARD_1880,DESITTER_1909}), but they are now replaced by purely numerical models, mainly used for ephemerides purposes (e.g., \citealp{LAINEY-etal_2004,LAINEY-etal_2004b}). Such models are extremely accurate but very computationally demanding. Even though some authors do adopt a purely numerical approach for moderately long timescales \citep{MUSOTTO-etal_2002}, the capabilities of such simulations remain way below the billions of years required by our study, especially when it comes to drawing a statistical picture of a chaotic event. Moreover, due to the chaotic nature of the dynamics and the finite-precision arithmetic of computers, it is not possible to obtain a precise orbital solution after a few thousand years. We must instead focus on the essential elements of the dynamics, which is the purpose of secular (i.e., averaged) theories.

\cite{LARI_2018} recently presented an averaged model that was shown to describe the orbital dynamics of the Galilean satellites with unprecedented precision, while keeping the advantage of being much faster than direct numerical integration; the model also includes tidal dissipation. This model is therefore an excellent starting point for our study, even though it requires some rearrangements: mainly the introduction of the suitable resonant terms.


The paper is structured as follows. In Sect.~\ref{sec:model}, we introduce the dynamical model used to integrate the motions of the Galilean satellites. In Sect.~\ref{sec:evol}, we describe our numerical experiments and analyze the outcomes of the simulations. In Sect.~\ref{sec:discuss}, we discuss the stability of the Laplace resonance and the variety of mean-motion resonances in which Callisto can be captured. In Sect.~\ref{sec:dtides}, we examine the robustness of our findings in view of the modeling of energy dissipation. Finally, we summarize our results in Sect.~\ref{sec:concl}.
   

\section{Dynamical model}\label{sec:model}
For the purpose of the present study, several improvements have been made to the model of \cite{LARI_2018}. Firstly, no Laplace coefficient is kept constant through time, and the equations of motion now include the partial derivatives of all Laplace coefficients. This ensures the validity of the model even if the ratios of semi-major axes vary substantially. Secondly, the orbit and obliquity of Jupiter are now allowed to vary with time according to a predefined solution. Using an appropriate evolution, the model is therefore valid over billions of years. Finally, the solar terms have been developed in Legendre polynomials and the expansion over the inclination of the Sun has been suppressed. This way, the solar contribution is more accurate (it is valid for any value of the obliquity of the planet), and numerous Laplace coefficients are no longer needed, allowing us to speed up the computations.

We also improved the implementation of the model. In particular, the integration coordinates have been changed, making the program more versatile, and a new algorithm has been implemented to compute the Laplace coefficients and their derivatives, making use of the Chebyshev interpolation. It is faster than before and accurate to machine precision. This way, we are assured that no numerical error other than round-off can add up to the truncation level inherent to the dynamical model. Below, we reiterate  the basic components of the model of \cite{LARI_2018} and highlight its modifications.

\subsection{Hamiltonian function}\label{sec:HamFunc}
The Hamiltonian function describing the long-term orbital dynamics of the Galilean satellites can be written
\begin{equation}\label{eq:H}
   \mathcal{H} = \mathcal{H}_0 + \varepsilon\mathcal{H}_1 \,,
\end{equation}
where the unperturbed part is a sum of Keplerian Hamiltonian functions:
\begin{equation}
   \mathcal{H}_0 = -\sum_{i=1}^N \frac{\mathcal{G}m_0m_i}{2a_i} \,,
\end{equation}
and the perturbation can be decomposed into
\begin{equation}\label{eq:eH1}
   \varepsilon\mathcal{H}_1 = \mathcal{H}_\text{J} + \mathcal{H}_\text{M} + \mathcal{H}_\odot + \mathcal{H}_\text{I} \,.
\end{equation}
In these expressions, the index $i$ runs over all satellites ($N=4$). $\mathcal{G}$ is the gravitational constant, $m_i$ and $a_i$ are the mass and the semi-major axis of the $i$th satellite, and $m_0$ is the mass of Jupiter. A parameter $\varepsilon\ll 1$ is used here to stress that $\varepsilon\mathcal{H}_1$ is small with respect to $\mathcal{H}_0$ (the explicit small parameters of each part of $\varepsilon\mathcal{H}_1$ are given below). We choose an equatorial reference frame, with the third axis oriented along the spin of Jupiter and the first axis directed towards its equinox (i.e., towards the ascending node of the Sun as seen in a Jovicentric reference frame).

The term $\mathcal{H}_\mathrm{J}$ in Eq.~\eqref{eq:eH1} is due to the nonsphericity of Jupiter. We consider that Jupiter has rotational and north--south symmetries, which is very close to reality \citep{IESS-etal_2018,SERRA-etal_2019}, and we use $R_\mathrm{J}$ to denote its equatorial radius. Up to second order in the eccentricity and inclination of the satellites, and up to fourth order in the ratio $R_\mathrm{J}/a_i$, the Hamiltonian $\mathcal{H}_\mathrm{J}$ can be written
\begin{equation}
   \begin{aligned}
      \mathcal{H}_\text{J} = \sum_{i=1}^N\frac{\mathcal{G}m_0m_i}{a_i}&\Bigg[J_2\left(\frac{R_\text{J}}{a_i}\right)^2\frac{1}{4}\Big(-2 - 3e_i^2 + 12s_i^2\Big) \\
      &+ J_4\left(\frac{R_\text{J}}{a_i}\right)^4\frac{3}{8}\Big(1 + 5e_i^2 - 20s_i^2\Big) \Bigg] \,,
   \end{aligned}
\end{equation}
where $J_2$ and $J_4$ are the zonal gravity harmonics of Jupiter, $e_i$ is the eccentricity of the $i$th satellite, $I_i$ its inclination, and $s_i\equiv\sin(I_i/2)$.

The term $\mathcal{H}_\mathrm{M}$ in Eq.~\eqref{eq:eH1} is due to the mutual gravitational attraction between the satellites. It can be further decomposed into a secular and a resonant part:
\begin{equation}
   \mathcal{H}_\text{M} = \mathcal{H}_\text{M}^{(\text{sec})} + \mathcal{H}_\text{M}^{(\text{res})} \,.
\end{equation}
Up to second order in the eccentricity and inclination of the satellites, the secular part can be written
\begin{equation}
   \begin{aligned}
      \mathcal{H}_\text{M}^{(\text{sec})} = -\sum_{1\leqslant i<j\leqslant N} \frac{\mathcal{G}m_im_j}{a_j}\Bigg(&
      f_1 + f_2(e_i^2+e_j^2) - \frac{1}{2}f_{14}(s_i^2+s_j^2) \\
      &+ f_{10}\ \ e_ie_j\ \ \cos(\varpi_j-\varpi_i) \\
      &+ f_{14}\ \ s_is_j\ \ \cos(\Omega_j-\Omega_i) \Bigg)\,,
   \end{aligned}
\end{equation}
where $\varpi_i$ is the longitude of perihelion of the $i$th satellite, $\Omega_i$ is its longitude of ascending node, and the $f_k$ functions are combinations of Laplace coefficients that depend on $a_i/a_j<1$ (see e.g., \citealp{MURRAY-DERMOTT_2000}). While the three inner satellites drift outwards due to tidal dissipation, the first low-order mean-motion resonance reached involving Callisto and another Galilean satellite is the 2:1 resonance with Ganymede. This means that after some time of evolution, the corresponding harmonics cannot be considered as fast angles (contrary to evolution close to the present time considered by e.g., \citealp{LARI_2018}). Accordingly, up to second order in the eccentricity and inclination of the satellites, the resonant part of the averaged mutual perturbations is:
\begin{equation}\label{eq:HMres}
   \begin{aligned}
      \mathcal{H}_\text{M}^{(\text{res})} = \!\!\!\!\!\!\!\sum_{ij=(12,23,34)}\!\!\bigg[\frac{\beta_in_ia_i\,\beta_jn_ja_j}{m_0} && e_j & \cos(\lambda_i - 2\lambda_j + \varpi_j)\\
      -\frac{\mathcal{G}m_im_j}{a_j}\Bigg(
        f_{27}            && e_i    & \cos(2\lambda_j -  \lambda_i -  \varpi_i) \\
      + f_{31}            && e_j    & \cos(2\lambda_j -  \lambda_i -  \varpi_j) \\
      + f_{45}            && e_i^2  & \cos(4\lambda_j - 2\lambda_i - 2\varpi_i) \\
      + f_{53}            && e_j^2  & \cos(4\lambda_j - 2\lambda_i - 2\varpi_j) \\
      + f_{49}            && e_ie_j & \cos(4\lambda_j - 2\lambda_i - \varpi_i - \varpi_j) \\
      - \frac{1}{2}f_{62} && s_i^2  & \cos(4\lambda_j - 2\lambda_i - 2\Omega_i) \\
      - \frac{1}{2}f_{62} && s_j^2  & \cos(4\lambda_j - 2\lambda_i - 2\Omega_j) \\
      + f_{62}            && s_is_j & \cos(4\lambda_j - 2\lambda_i - \Omega_i - \Omega_j)
      \Bigg)\Bigg]\,,
   \end{aligned}
\end{equation}
where $\lambda_i$ is the mean longitude of the $i$th satellite, $\beta_i=m_0m_i/(m_0+m_i)$, and $n_i^2a_i^3=\mu_i=\mathcal{G}(m_0+m_i)$. The first term is the indirect part of the perturbation (see Appendix~\ref{asec:ham}), whose expression was not explicitly given by \cite{LARI_2018}. The terms with indexes $(i,j)=(3,4)$ correspond to the 2:1 resonance between Ganymede and Callisto, which is absent from \cite{LARI_2018}.

The term $\mathcal{H}_\odot$ in Eq.~\eqref{eq:eH1} is due to the gravitational attraction of the Sun. Since the Sun is much farther away from Jupiter than the satellites, it is convenient to expand its perturbation in Legendre polynomials. This way, we can avoid any expansion with respect to the Sun's inclination \citep{LASKAR-BOUE_2010}, meaning that the expression remains valid for any value of the obliquity of the planet considered. We write $a_\odot$ the semi-major axis of the Jovicentric orbit of the Sun. Up to fourth order in the ratio $a_i/a_\odot$, the perturbation from the Sun can be written
\begin{equation}\label{eq:Hsun}
   \begin{aligned}
      \mathcal{H}_\odot = \sum_{i=1}^N\frac{\mathcal{G}m_\odot m_i}{a_\odot}\Bigg[ \left(\frac{a_i}{a_\odot}\right)^2&\Bigg(
      {C}_1^\odot + {C}_2^\odot(e_i^2 - 4s_i^2) \\
      &+ {C}_3^\odot s_i^2\cos(2\Omega_i) + {C}_4^\odot s_i^2\sin(2\Omega_i) \\
      &+ \frac{5}{4}{C}_3^\odot e_i^2\cos(2\varpi_i) + \frac{5}{4}{C}_4^\odot e_i^2\sin(2\varpi_i) \\
      &+ {C}_5^\odot s_i\cos\Omega_i + {C}_6^\odot s_i\sin\Omega_i
      \Bigg) \\
      +\left(\frac{a_i}{a_\odot}\right)^3&\Bigg( {C}_7^\odot e_i\cos\varpi_i + {C}_8^\odot e_i\sin\varpi_i \Bigg) \\
      +\left(\frac{a_i}{a_\odot}\right)^4&{C}_9^\odot \Bigg] \,,
   \end{aligned}
\end{equation}
where the coefficients $C_1^\odot$ to $C_9^\odot$ are known functions of the time that only depend on the orbital elements of the Sun, including its mean longitude (see Appendix~\ref{asec:ham}). For each degree in $a_i/a_\odot$, the order of the expansion in $e_i$, $s_i$, and the Sun's eccentricity has been adjusted in such a way that all neglected terms have the same order of magnitude.

The term $\mathcal{H}_\mathrm{I}$ in Eq.~\eqref{eq:eH1} is due to inertial forces. This perturbation was not present in \cite{LARI_2018} because the obliquity of Jupiter and its orbit around the Sun were considered fixed. If the obliquity and orbit of Jupiter are considered to vary with time (and they do vary over long timescales), the reference system attached to Jupiter's equator is not inertial anymore. This means that additional accelerations apply to the satellites, like the centrifugal or Coriolis terms. As shown in Appendix~\ref{asec:ham}, the noninertial nature of the reference frame can be taken into account by redefining the canonical coordinates used, leading to the following expression:
\begin{equation}
   \begin{aligned}
      \mathcal{H}_\text{I} = \sum_{i=1}^N\beta_i\sqrt{\mu_ia_i}\Big[&- \Theta_z && \Big(1 - 2s_i^2 - \frac{e_i^2}{2}\Big) \\
      &+ 2\Theta_y && s_i \cos\Omega_i \\
      &- 2\Theta_x && s_i \sin\Omega_i
      \Big]\,,
   \end{aligned}
\end{equation}
where $\mathbf{\Theta}=(\Theta_x,\Theta_y,\Theta_z)^\mathrm{T}$ is the instantaneous rotation vector of our noninertial reference frame as measured in an inertial reference frame (here, the J2000 ecliptic and equinox). The explicit expression of $\mathbf{\Theta}$ in terms of the orbital elements and obliquity of Jupiter is given in Appendix~\ref{asec:QPS}. It is zero if the orbit and obliquity of Jupiter are fixed in time.

In order to express the equations of motion, we need to choose a set of canonical coordinates. We start from the modified Delaunay canonical coordinates:
\begin{equation}\label{eq:Delaunay}
   \left\{
   \begin{aligned}
      L_i &= \beta_i\sqrt{\mu_i a_i} \\
      G_i &= \beta_i\sqrt{\mu_i a}\bigg(1-\sqrt{1-e_i^2}\bigg) \\
      H_i &= \beta_i\sqrt{\mu_i a_i(1-e_i^2)}\bigg(1-\cos I_i\bigg)
   \end{aligned}
   \right.
   \text{and}\hspace{0.5cm}
   \left\{
   \begin{aligned}
      \ell_i &= \lambda_i \\
       g_i &= -\varpi_i \\
       h_i &= -\Omega_i,
   \end{aligned}
   \right.\,,
\end{equation}
where uppercase characters are the momenta, and lowercase characters are their conjugate angles\footnote{There is a typographical error for variable $H_i$ in \cite{LARI_2018}.}. Since our Hamiltonian function is truncated at second order in eccentricity and inclination, we use the following relations:
\begin{equation}
   e_i = \sqrt{\frac{2G_i}{L_i}} + \mathcal{O}(e_i^3)
   \,,\hspace{1cm}
   s_i = \sqrt{\frac{H_i}{2L_i}} + \mathcal{O}(e_i^2s_i)
   \,,
\end{equation}
and neglect the remainders. We get rid of the coordinate singularities at $e_i=0$ and $s_i = 0$ by the use of rectangular canonical coordinates:
\begin{equation}\label{eq:xyuv}
   \left\{
   \begin{aligned}
      x_i &= \sqrt{2G_i}\cos g_i \\
      u_i &= \sqrt{2H_i}\cos h_i \\
   \end{aligned}
   \right.
   \hspace{0.5cm}\text{and}\hspace{0.5cm}
   \left\{
   \begin{aligned}
      y_i &= \sqrt{2G_i}\sin g_i \\
      v_i &= \sqrt{2H_i}\sin h_i \\
   \end{aligned}
   \right.\,.
\end{equation}
Finally, we introduce the resonant canonical coordinates by replacing $L_i$ and $\ell_i$ with
\begin{equation}\label{eq:gam}
   \left\{
   \begin{aligned}
      \Gamma_1 &=  L_1 \\
      \Gamma_2 &= 2L_1 +  L_2 \\
      \Gamma_3 &= 4L_1 + 2L_2 +  L_3 \\
      \Gamma_4 &= 8L_1 + 4L_2 + 2L_3 + L_4 \\
   \end{aligned}
   \right.
   \text{and}\hspace{0.5cm}
   \left\{
   \begin{aligned}
      \gamma_1 &= \lambda_1 - 2\lambda_2 \\
      \gamma_2 &= \lambda_2 - 2\lambda_3 \\
      \gamma_3 &= \lambda_3 - 2\lambda_4 \\
      \gamma_4 &= \lambda_4 \\
   \end{aligned}
   \right.\,.
\end{equation}
The generic form of these coordinates makes it easy to add or remove one satellite for test purposes.

Since the Hamiltonian function has been averaged over short-period terms, it does not depend on $\gamma_4$. This makes $\Gamma_4$ a constant of motion in the conservative case. The other variables evolve according to Hamilton's equations. The total Hamiltonian function in Eq.~\eqref{eq:H} explicitly depends on time through the coefficients $C_1^\odot$ to $C_9^\odot$, and through the vector~$\mathbf{\Theta}$. Both are functions of the obliquity and orbit of Jupiter. The orbital evolution of Jupiter is taken from state-of-the-art secular theories \citep{LASKAR_1990} combined with the \texttt{INPOP17a} modern ephemerides\footnote{\texttt{https://www.imcce.fr/inpop/}}. A solution for the secular dynamics of Jupiter's spin-axis is obtained numerically. The resulting orbital and rotational solutions are put into the form of quasi-periodic series, allowing for extremely fast function evaluations. More details about how these solutions are built can be found in Appendix~\ref{asec:QPS}.

The use of an averaged model allows us to greatly speed up the numerical integrations. The accuracy of the numerical integration can be checked by monitoring the value of the Hamiltonian function in Eq.~\eqref{eq:H}, when considering a fixed orbit and obliquity for Jupiter and no dissipation. Using the numerical integrator of \cite{EVERHART_1985} refined using the tips given by \cite{REIN-SPIEGEL_2015}\footnote{The predictor--corrector iterations are stopped only when full convergence has been reached, and every step-size or convergence control is made using nondimensional quantities.}, we found that a constant step size of $11$~days is a good compromise (compared to a step size of less than one hour, which would be required in a nonaveraged model).

\subsection{Tidal dissipation}\label{subsec:tides}
Tides are differential gravitational forces acting on an extended body. Their main effect is to raise two tidal bulges along the direction between the body that generates them and the one that is exposed to their effects. This redistribution of mass induces an additional gravitational field around the deformed body, which is proportional to the Love number $k_2$ \citep{DARWIN_1880,LOVE_1909,KAULA_1964}. For realistic bodies, the response to the tidal perturbation is not immediate but has a time delay, which results in a shift of the tidal bulges of a certain angle $\delta$ (see e.g., \citealp{MACDONALD_1964,SINGER_1968,MIGNARD_1979}) accompanied by a loss of energy due to internal friction. Both $\delta$ and $k_2$ depend on the interior structure of the body and its rheology (e.g., \citealp{EFROIMSKY-MAKAROV_2013,BOUE-etal_2016,BOUE-etal_2019}). The angle $\delta$ is related to the quality factor $Q$ \citep{MACDONALD_1964}, which is the amount of orbital energy over the dissipated energy per orbit due to tidal friction. The smaller the value of $Q$, the larger the dissipation inside the tidally deformed body. The value of this parameter can go from tens to hundreds for terrestrial bodies and from thousands to millions for gas giants. For an overview of energy dissipation in the Solar System, see \cite{GOLDREICH-SOTER_1966}. More recently, \cite{FERRAZMELLO_2013} developed a new theory of dynamical tides based on a simple rheophysical model of the bodies. In this model, phase lags (and therefore $Q$) are not ad hoc quantities designed to model the delayed response to the tides, but they are determined from the solutions of the equations.

For the aim of this work, we are interested in the long-term dynamical effects of the tidal forces. From \cite{KAULA_1964} and \cite{PEALE-CASSEN_1978}, we know that for a couple formed by a planet and a synchronous satellite $i$, tides cause a secular variation of the satellite's semi-major axis $a_i$, eccentricity $e_i$, and inclination $I_i$ (see also \citealp{FERRAZMELLO-etal_2008,BOUE-EFROIMSKY_2019}). However, no inclination-type resonance enters into play for the Galilean satellites nowadays or during the resonant encounter with Callisto (this is verified in Sect.~\ref{sec:evol}.). For this reason, we can neglect the tidal effects on the orbital inclinations: if included, the dissipation would simply damp their already low values, making their contribution to the dynamics even more marginal. The variation of the semi-major axis and the eccentricity of a satellite due to the tidal dissipation can be described by the following formulas:
\begin{align}
   \label{dissa}
   \dot{a}_i &= \frac{2}{3}c_i\left(1-\left(7D_i-\frac{51}{4}\right)e_i^2\right)a_i\,, \\
   \label{disse}
   \dot{e}_i & =-\frac{1}{3}c_i\left(7D_i-\frac{19}{4}\right)e_i\,;
\end{align}
for anelastic tides, where, using the notation of \cite{MALHOTRA_1991},
\begin{equation}\label{eq:dissmalho}
   \begin{aligned}
      c_i&=\frac{9}{2}\left(\frac{k_2}{Q}\right)_{0,i}\frac{m_i}{m_0}\left(\frac{R_\mathrm{J}}{a_i}\right)^5n_i\,,\\
      D_i&=\left(\frac{k_2}{Q}\right)_i\left(\frac{Q}{k_2}\right)_{0,i}\left(\frac{R_i}{R_\mathrm{J}}\right)^5\left(\frac{m_0}{m_i}\right)^2\,;
   \end{aligned}
\end{equation}
with $R_i$ being the  radius of the satellite. The ratios $(k_2/Q)_i$ and $(k_2/Q)_{0,i}$ are the dissipative parameter of the $i$th satellite and the dissipative parameter of the planet at the orbital frequency of the $i$th satellite, respectively. Indeed, from \cite{FULLER-etal_2016} and \cite{LAINEY-etal_2017}, we know that tidal dissipation within a planet can strongly depend on the satellite that raises the tides.

In the case of the couple Jupiter--Io, the most reliable estimate of the dissipative parameters was obtained by \cite{LAINEY-etal_2009}, who fitted a complete numerical model to astrometric observations taken from 1891 to 2007. The orbit determination revealed a strong energy dissipation within Io and Jupiter, with values:
\begin{equation}\label{eq:dissipIo}
   \begin{aligned}
      (k_2/Q)_1 &= (1.5\pm 0.3)\times 10^{-2}\,,\\
      (k_2/Q)_{0,1} &= (1.1\pm 0.2)\times 10^{-5}\,. \\
   \end{aligned}
\end{equation}
Solely the dissipation in the couple Jupiter--Io has been estimated so far because tidal forces are larger for satellites that are closer to the planet.

As Io, Europa, and Ganymede are locked in mean-motion resonance, they adiabatically follow the slow drift of the resonance center due to the dissipation (this is verified in Sect.~\ref{sec:evol}). This means that their ratios of semi-major axes remain approximately constant during the evolution, such that the 4:2:1 commensurability is maintained, and therefore we always have
\begin{equation}\label{eq:resadot}
   \frac{\dot{a}_2}{\dot{a}_1}\bigg|_\text{res} \approx \frac{a_2}{a_1} \approx 1.6
   \hspace{0.5cm}\text{and}\hspace{0.5cm}
   \frac{\dot{a}_3}{\dot{a}_1}\bigg|_\text{res} \approx \frac{a_3}{a_1} \approx 2.5 \,.
\end{equation}
Because of the mean-motion resonance, we do not expect values for $(k_2/Q)_{0,2}$ and $(k_2/Q)_{0,3}$ much different from the one observed for Io. Therefore, a high upper bound for the drift of the semi-major axes of Europa and Ganymede due to their intrinsic tidal dissipation can be obtained by assuming that they have the same dissipation parameters as Io,  or similar. From Eq.~\eqref{dissa}, and assuming the same values as in Eq.~\eqref{eq:dissipIo}, the effect of the respective tides of Europa and Ganymede on their semi-major axes is
\begin{equation}
   \frac{\dot{a}_2}{\dot{a}_1}\bigg|_\text{dis} \approx 0.04
   \hspace{0.5cm}\text{and}\hspace{0.5cm}
   \frac{\dot{a}_3}{\dot{a}_1}\bigg|_\text{dis} \approx 0.01 \,.
\end{equation}
These drifts are much smaller than the ones imposed by the resonant link (compare with Eq.~\ref{eq:resadot}). Their contribution is even smaller than the error bars coming from the uncertainty of $(k_2/Q)_{0,1}$ and $(k_2/Q)_{1}$. Consequently, we can safely neglect the contribution of Europa and Ganymede to the energy dissipation, and only consider the contribution from Io.

The dissipation parameters of Callisto are even less constrained, and Callisto is currently not involved in any mean-motion resonance. However, considering again the same dissipation parameters as Io, we obtain
\begin{equation}
   \frac{\dot{a}_4}{\dot{a}_1}\bigg|_\text{dis} \approx 0.0003 \,.
\end{equation}
This very small ratio shows that a dramatically high (and improbable) value of $(k_2/Q)_{0,4}$ would be needed in order for Callisto to reach a migration rate comparable to those of Io, Europa, and Ganymede. In other words, Callisto can be considered as almost stationary with respect to the migration rates of the other moons. Consequently, we also neglect its contribution to the energy dissipation.

As seen above, the approximation of only considering the tidal dissipation generated between Io and Jupiter is justified by the many orders of magnitude that separate its level for Io and for the other Galilean satellites. Moreover, no estimate of the dissipation parameters has been obtained yet from observations for satellites other than Io (neither from astrometry nor from space missions). Therefore, instead of exploring a range of ad hoc values for these unknown parameters, we prefer to neglect them, taking advantage of their small impact on the dynamics. This approach has already been used for instance by \cite{DEIENNO-etal_2014}; it reduces the parameter space to explore, helping us to develop a clear understanding of the simplified dynamics before any investigation of the extra level of detail that would be provided by more realistic models. In any case, underestimating the tidal dissipation in the system would mainly change the timescale of the long-term evolution of the satellites, and not its qualitative behavior (see Sect.~\ref{sec:dtides} for more details).

Nevertheless, it should be noted that neglecting the energy dissipation between Jupiter and Europa, Ganymede, and Callisto also suppresses the direct damping of their eccentricities presented in Eq.~\eqref{disse}. The eccentricity of Europa is almost entirely forced today ($e_2^\mathrm{forced}\approx 0.010$ and $e_2^\mathrm{free}\approx 0.000$, see \citealp{SINCLAIR_1975}) while the eccentricity of Ganymede only contains a small free component to be damped ($e_3^\mathrm{forced}\approx 0.0010$ and $e_3^\mathrm{free}\approx 0.0005$, see \citealp{SINCLAIR_1975,NOYELLES-VIENNE_2007}). Callisto is the only Galilean satellite currently possessing a substantial free eccentricity ($e_4^\mathrm{forced}\approx 0.0002$ and $e_4^\mathrm{free}\approx 0.0071$, see \citealp{NOYELLES-VIENNE_2007}), but considering its large distance from Jupiter, its damping for realistic values of the dissipation parameter $(k_2/Q)_4$ is quite small even during the gigayear timescale spanned by our numerical simulations (decrease of about $0.0010$ in $1.5$ Gyrs assuming $(k_2/Q)_4=10^{-3}$).

Since they have been directly estimated from observations, the values of the dissipative parameters given in Eq.~\eqref{eq:dissipIo} can be considered as instantaneous quantities. More generally, it is well known that each quality factor $Q$ is a function of the tidal frequency $\chi$, which is forced to change because of the migration of the satellite. \cite{MIGNARD_1979} proposed that $Q = \Delta t\ \chi^{-1}$, assuming a constant tidal time-lag $\Delta t$. More recently, \cite{FERRAZMELLO_2013} showed that, in the pseudo-synchronous approximation,  $Q\propto\chi^{-1}$ for inviscid bodies, while $Q\propto\chi$ for high-viscosity bodies. Between these extreme cases, \cite{EFROIMSKY-LAINEY_2007} noted that a constant value of $Q$ is more realistic according to planetary interior models, at least for terrestrial bodies. Moreover, these latter authors reiterated the fact that $Q$ depends on the temperature of the body as well, and they presented a more general model in which
\begin{equation}\label{eq:Qbeta}
   Q=(\,\epsilon\chi\,)\,^\beta\,,
\end{equation}
where $\epsilon$ is a function of the temperature of the body and $\beta$ is a parameter encompassing the response of the dissipation to the tidal frequency. For terrestrial bodies, \cite{EFROIMSKY-LAINEY_2007}  predicted values of $\beta$ ranging from $0.2$ to $0.4$. Furthermore,  \cite{OJAKANGAS-STEVENSON_1986} and \cite{HUSSMANN-SPOHN_2004} showed that Io could suffer from temperature variation cycles with a period of about $100$~Myr. This would periodically change its quality factor $Q$, as well as its forced eccentricity. From these results, it appears that the development of a realistic frequency-dependent model of tidal dissipation would require the understanding of internal processes taking place inside the Galilean satellites. This is far beyond the scope of the present article.

Fortunately, at the level of generality required by our exploratory study, simple arguments show that there is no need for such refined models. It can be easily shown that the migration of Io needed for Ganymede to reach the 2:1 mean-motion resonance with Callisto (so that all four satellites are in a 2:1 chain of period ratios) only amounts to changing $a_1$ by a factor $1.1$. The frequency of the tides raised on Jupiter by Io is equal to:
\begin{equation}
   \chi_{0,1} = 2(w - n_1)\,,
\end{equation}
where $w$ is the spin velocity of Jupiter. By multiplying $a_1$ by a factor $1.1$, we obtain that $\chi_{0,1}$ changes by a factor of about $1.04$ only. Likewise, the frequency of the tides raised on Io by Jupiter is equal to $\chi_1=n_1$, which changes by a factor of about $0.87$ when $a_1$ is multiplied by $1.1$. Consequently, even using the extreme values of $\beta=\pm 1$ for the frequency dependence of $Q$ (see Eq.~\ref{eq:Qbeta}), we find that in both cases, the variations of $Q$ remain smaller than the uncertainties of its current value, quoted in Eq.~\eqref{eq:dissipIo}. As can be seen in Sect.~\ref{sec:evol}, this remains true over the whole duration of our numerical integrations. For this reason, the use of a refined frequency-dependent tidal model appears quite unnecessary in the context of our study. Therefore, we choose to use constant values for the dissipative parameters $(k_2/Q)_1$ and $(k_2/Q)_{0,1}$ obtained from their estimates given in Eq.~\eqref{eq:dissipIo}.

Because of the adiabatic nature of the tidal dissipation, accounting for the time dependency of $k_2/Q$ would mostly change the timing of the resonant encounter with Callisto and hardly its topological features. Only extremely different dissipation scenarios could make the orbits vary in such a way as to transform the topology of the encounter. This crucial point is further discussed in Sect.~\ref{sec:dtides}. In particular, extreme dissipation variations could be produced within Jupiter if during its migration one of the satellites reaches a resonance with an oscillation mode of the interior of Jupiter. As explained by \cite{FULLER-etal_2016}, the frequencies of such oscillation modes are not fixed with time but gradually shift because of the evolution of Jupiter's internal structure (e.g., due to its cooling). A resonance-driven dissipation peak could therefore reach the location of a satellite and then drag it along, as the satellite is forced to migrate faster following the shift of the peak. It is unknown whether this mechanism has already triggered for Io (in which case the corresponding quality factor should remain constant, imposed by the resonance), or whether it will be triggered in the future (in which case the dissipation will increase). Given our current level of ignorance, it seems reasonable to continue using constant dissipative parameters, at least in the context of this preliminary investigation of the future dynamics of the Galilean satellites.

Following \cite{MALHOTRA_1991} and \cite{LARI_2018}, we model dissipative effects as an adiabatic process. Indeed, even though~\eqref{dissa} and~\eqref{disse} are not conservative and cannot be derived from the Hamiltonian function described in Sect.~\ref{sec:HamFunc}, their effects are very small and act on very large time spans (millions of years), well separated from the characteristic resonant (a few years) and secular timescales (hundreds of years) of the motion of the Galilean satellites. This means that in the vicinity of any time $t$, the conservative dynamical system from Eq.~\eqref{eq:H} is valid, but that on greater timescales, the eccentricity and semi-major axis of Io follow the trends given in Eqs.~\eqref{dissa} and \eqref{disse}. Therefore, these trends can simply be added to the dynamical equations after having converted them in terms of the canonical coordinates given in Eqs.~\eqref{eq:xyuv} and \eqref{eq:gam}. 

In reality, the dissipative effects described above are so small (i.e., so well adiabatic) that we can even use a multiplying factor $\alpha$ to the dissipative parameters, following the approach of \cite{MALHOTRA_1991} and \cite{SHOWMAN-MALHOTRA_1997}. Using a dissipative parameter $\alpha$ times larger implies that the migration of the satellites is $\alpha$ times faster, which drastically reduces the computation time. This linear acceleration law can be proven by linearizing the small semi-major drift resulting from the tidal dissipation (see Eq.~\ref{dissa}). This method is valid as long as the accelerated energy dissipation remains adiabatic with respect to the conservative part of the dynamics. The choice of a suitable value for $\alpha$ is therefore critical. By examining the characteristic timescales of the dynamics of the Galilean satellites, \cite{MALHOTRA_1991} set an upper limit for $\alpha$ below which the evolution is not distorted by the artificial acceleration. Based on this result, \cite{MALHOTRA_1991} and \cite{SHOWMAN-MALHOTRA_1997} used an acceleration factor of about $10^3$ in their simulations. Here, we make a more conservative choice and set $\alpha$ to $10^2$. In the following sections, we give the results as a function of the physical time, which is obtained as the integration time multiplied by $\alpha$. Therefore, the gigayear scale in the figures of Sect.~\ref{sec:evol} represents $10$~Myr of actual integration time. The validity of this acceleration method and the quality of the adiabatic approximation is checked and further investigated in Sect.~\ref{sec:dtides}.

\subsection{Initial conditions}\label{sec:condinit}
We start our integration at time J2000. We use the same method as \cite{LARI_2018} in order to build suitable initial conditions for the semi-secular model: we filter the series of orbital elements taken from the \texttt{Jup310} ephemerides\footnote{\texttt{https://naif.jpl.nasa.gov/pub/naif/generic\_kernels}}, removing the short-period harmonics. As shown by \cite{LARI_2018}, integrations with our model for $100$~years (about the time that ephemerides cover) are in very good agreement with the filtered series of \texttt{Jup310}. This means that this model very accurately reproduces the resonant and secular dynamics of the Galilean satellites. The system is then propagated forward for billions of years. The values of the parameters and of the initial conditions used in this article are given in Table~\ref{tab:par}.

\begin{table}[t]
        \caption{Physical and orbital parameters used in this article.}
        \label{tab:par}
        \centering
        \begin{tabular}{l r r} 
                \hline\hline
                Par. & Value $(R_\text{J}, m_0, \text{yr})$ & Value $(\text{km}, \text{kg}, \text{s})$ \\
                \hline
                $\mathcal{G}$&  346166894.5504444  & $6.67259\times 10^{-20}$   \\
                yr           &                  1  & 31557600   \\
                $m_\odot$    &  1047.571735402983  & $1.98893133312\times 10^{30}$   \\
                $m_0$        &                  1  & $1.89861110786\times 10^{27}$   \\
                $m_1$        &  0.000047044621535  & $8.93194410114\times 10^{22}$   \\
                $m_2$        &  0.000025280745501  & $4.79983042232\times 10^{22}$   \\
                $m_3$        &  0.000078049574517  & $1.48185789142\times 10^{23}$   \\
                $m_4$        &  0.000056669717201  & $1.07593754557\times 10^{23}$   \\
                $R_\text{J}$ &                  1  & 71398\phantom{.0}   \\
                $R_1$        &  0.025513319700832  & 1821.6   \\
                $J_2$        &  0.014735\phantom{000000000}  &   * \\
                $J_4$        & -0.0005888\phantom{00000000}  &   * \\
                $a_1$        &  5.91907361630506   &  422610.018056949   \\
                $a_2$        &  9.41465350340912   &  672187.430836404   \\
                $a_3$        &  15.01570120737033  & 1072091.034803830   \\
                $a_4$        &  26.41170891766961  & 1885743.193303770   \\
                $e_1$        &  0.004139765215464  &    *    \\
                $e_2$        &  0.009526378335510  &    *    \\
                $e_3$        &  0.001453747714343  &    *    \\
                $e_4$        &  0.007404398442897  &    *    \\
                $I_1$        &  0.000661204620550  &    *    \\ 
                $I_2$        &  0.008068640006861  &    *    \\
                $I_3$        &  0.003605612584037  &    *    \\
                $I_4$        &  0.003482321688065  &    *    \\
                $\omega_1$   &  2.884294432504667  &    *    \\
                $\omega_2$   &  0.853911845165254  &    *    \\
                $\omega_3$   &  4.083549935722539  &    *    \\
                $\omega_4$   &  3.431308774505764  &    *    \\
                $\Omega_1$   &  1.820488369209967  &    *    \\
                $\Omega_2$   &  0.718573110624886  &    *    \\
                $\Omega_3$   &  5.129820134251226  &    *    \\
                $\Omega_4$   &  0.339041833619144  &    *    \\
                $\gamma_1$   &  1.571055378146310  &    *    \\
                $\gamma_2$   &  4.709594574392012  &    *    \\
                $\gamma_3$   &  3.470563159062861  &    *    \\
                \hline
        \end{tabular}
        \tablefoot{The values are given in two different systems of units (columns). The angles are all in radians. Physical parameters are taken from the L3 ephemerides \citep{LAINEY-etal_2009}, and the orbital elements are given at time J2000; they are drawn from the averaged Jovicentric canonical coordinates (see Appendix~\ref{asec:ham}), computed by filtering the numerical ephemerides (see Sect.~\ref{sec:condinit}). The asterisk signifies that the value is the same in both  systems of units. The number of digits are not representative of the uncertainties; numerous digits are given for repeatability.}
\end{table}


\section{Long-term evolution}\label{sec:evol}

\begin{figure}
   \centering
   \includegraphics[scale=0.7]{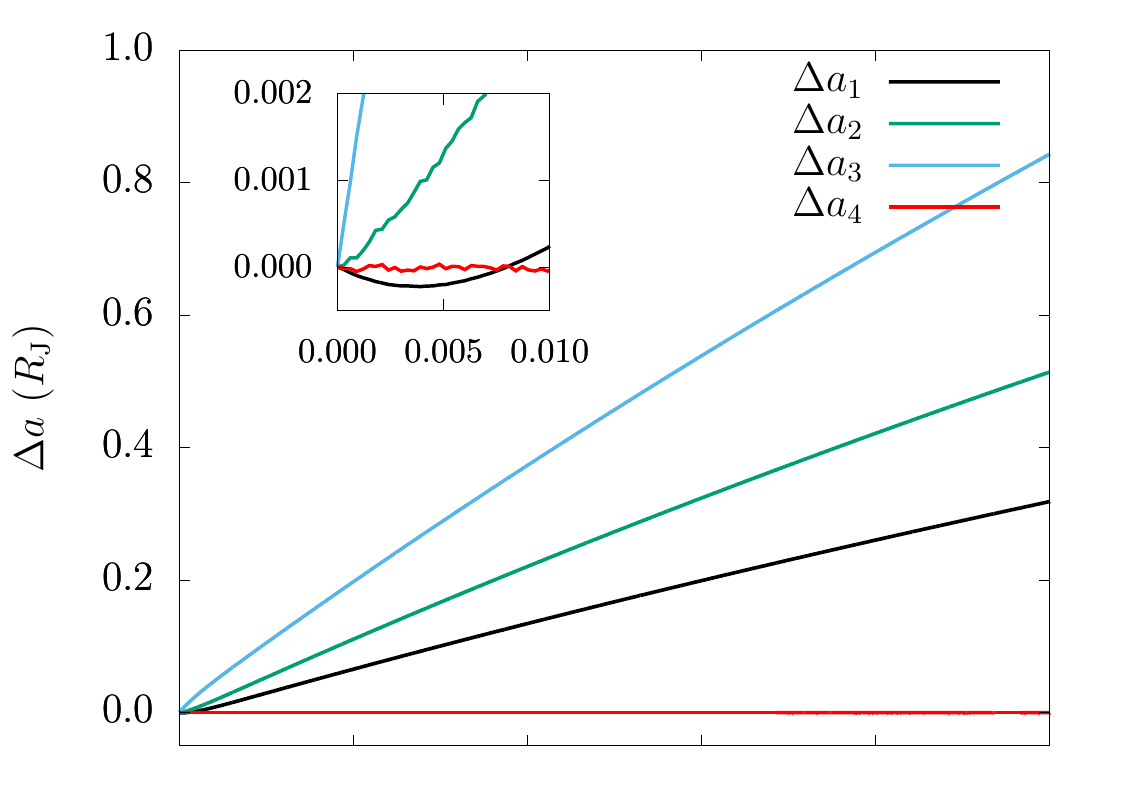}\\
   \includegraphics[scale=0.7]{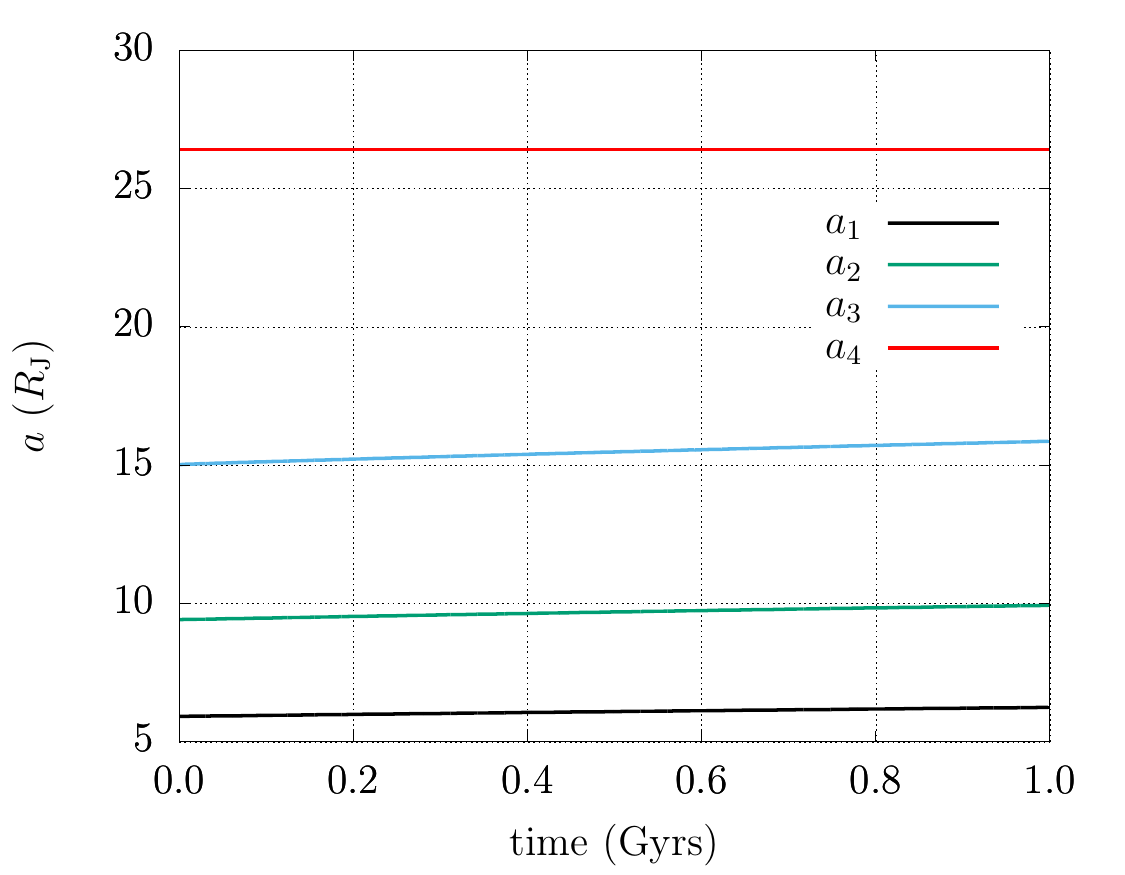}
   \caption{Variation of the satellites' semi-major axes ($\Delta a$ and $a$) in the first phase of the evolution. Due to the Laplace resonance, the tidal dissipation is distributed among Io, Europa, and Ganymede. As shown in the zoom-in view, Io initially migrates inward and then outward like Europa and Ganymede. Callisto does not have any secular trend.}
   \label{fig:awzoom}
\end{figure}

\begin{figure}
   \centering
   \includegraphics[scale=0.7]{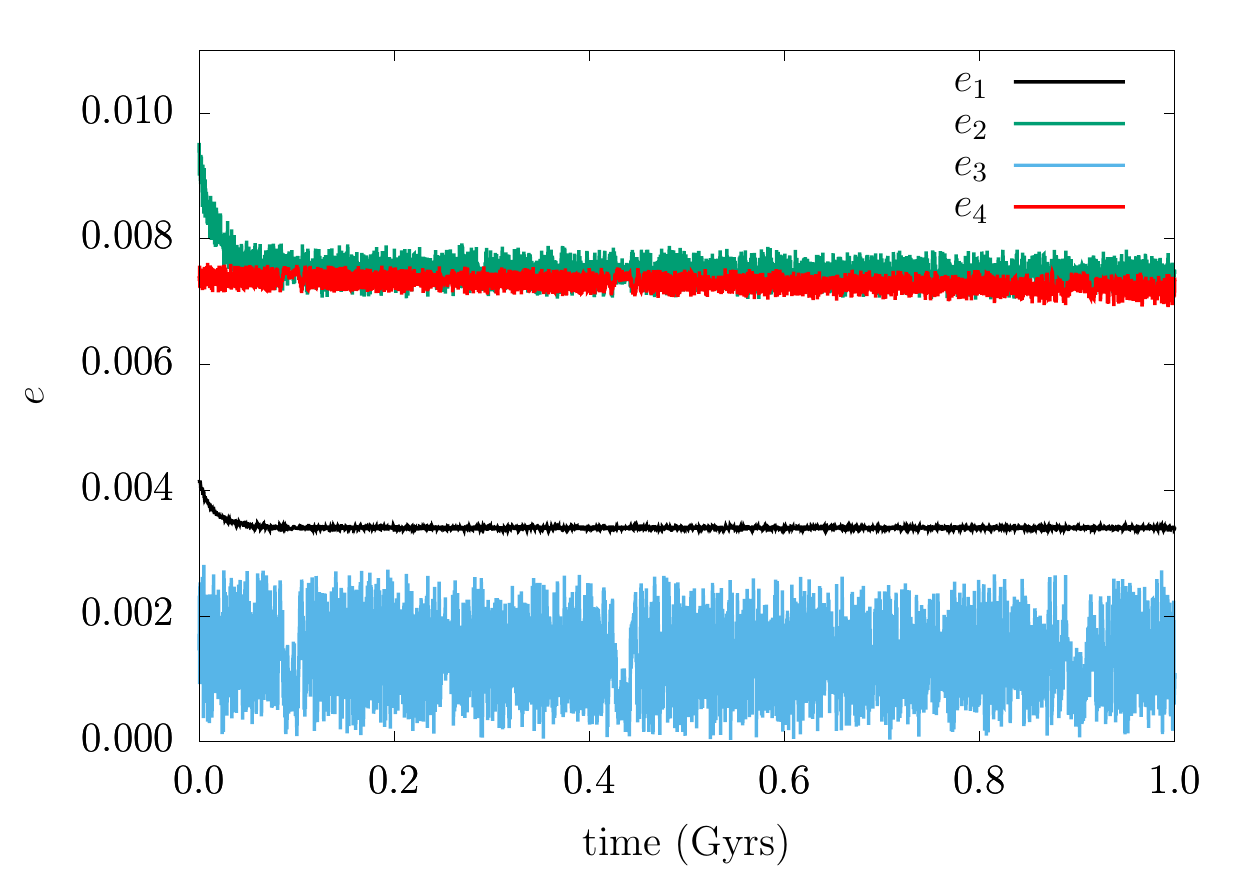}
   \caption{Variation of the satellites' eccentricities in the first phase of the evolution. Io and Europa's eccentricities initially decrease, and then, when $a_2/a_1$ remains almost constant, they stabilize to new values.}
   \label{fig:ewzoom}
\end{figure}

The current configuration of the system consists in a chain of two 2:1 mean-motion resonances in the couples Io--Europa and Europa--Ganymede. From \cite{LAINEY-etal_2009}, we know that at present Io is moving toward Jupiter, while Europa and Ganymede move away from the planet. However, on a long timescale, the tidal dissipation results in an outward migration for all the satellites. Indeed, as shown in \figurename~\ref{fig:awzoom}, after about $4$~Myr Io stops its inward migration and starts migrating outwards. This inversion is not due to a change of sign in Eq.~\eqref{dissa}: the slow trend imposed by the tidal dissipation always remains positive. Instead, this change of direction is due to the fact that $a_1$ decreases and $a_2$ increases, meaning that the ratio $a_2/a_1$ changes rapidly (while remaining close to the value quoted in Eq.~\ref{eq:resadot}). This shift of the resonance center between Io and Europa induces a variation of the forced values of their eccentricities (see \figurename~\ref{fig:ewzoom}). Since the eccentricity of Io decreases, dissipation in Jupiter gains importance against the one within Io (see Eq.~\ref{dissa}). This provides more energy to the orbit of Io and makes all three semi-major axes increase.

This current, surprising behavior of the Galilean satellites may conceal some clues about the origin of the Laplace resonance. Today, the mean-motion relations $n_1-2n_2$ and $n_2-2n_3$ increase because the satellites migrate in different directions. This increase has been reported for instance by \cite{LAINEY-etal_2009}. As explained in Sect.~\ref{sec:intro}, it could seem to favor a primordial origin of the Laplace resonance, since the satellites would now be evolving away from deep resonance. However, this ``decay'' of the resonance, put forward for instance by \cite{PEALE-LEE_2002}, stops after a relatively short amount of time, once an equilibrium is reached between the resonant dynamics and the dissipative effects. The Laplace resonance is fully preserved. Therefore, this temporary behavior could hardly be related to a global trend pushing the system away from a primordial hypothetical state. On the contrary, it could indicate that the Laplace resonance is quite new, since it has not  yet reached an equilibrium configuration. More probably, this transition could be due to cyclic variations of the dissipation parameters, periodically forcing the Laplace resonance to slightly resettle at the new equilibrium configuration. As detailed in Sect.~\ref{subsec:tides}, such variations could be due to internal processes of the planet (e.g., \citealp{BURKART-etal_2014}), and/or of the satellite (e.g., \citealp{OJAKANGAS-STEVENSON_1986,HUSSMANN-SPOHN_2004}). However, as pointed out by \cite{FULLER-etal_2016}, we expect in the long run to observe an outward migration of all three Galilean satellites, as shown in \figurename~\ref{fig:awzoom}. In case of cyclic variations, the constant dissipation parameters used in our model should therefore be interpreted as mean values, representative of the global trend of the system.

Using the values of the dissipative parameters from Eq.~\eqref{eq:dissipIo}, we obtain that for about $1.4$~Gyr from today the evolution is very stable: all the current resonances are preserved, and small differences in the initial conditions do not change the qualitative behavior of the resonant angles nor the timescale of the migration. This proves the stability of the Laplace resonance under the action of tidal dissipation over very long timescales.

However, after $1.4$~Gyr,  as Ganymede approaches the 2:1 mean-motion resonance with Callisto, chaotic effects show up: orbital elements suddenly change because of the exit from mean-motion resonances or the capture into new ones. From this point on, a small change in the variables (or in the model) results in a completely different evolution of the system. For this reason, we adopt a statistical approach to study the outcome of the resonant encounter. Since Callisto is initially out of any mean-motion resonance, its mean longitude (contained in the variable $\gamma_3$) at a given time can be considered as random with respect to the longitude of any other satellite in the system. As a result, a tiny error in the initial conditions of the satellites (or in the dynamical model) is transformed after $1.4$~Gyr into a uniform distribution of $\gamma_3$ in $[0,2\pi)$. Hence, starting from the coordinates at $1.4$~Gyr obtained from our nominal propagation, we generate a list of new initial conditions, in which $\gamma_3$ is sampled in the whole interval $[0,2\pi)$ while the other variables are kept the same. We use a sampling step of about $0.01$ radians so that the total number of simulations is $628$.

\begin{figure}
   \centering
   \includegraphics[scale=0.72]{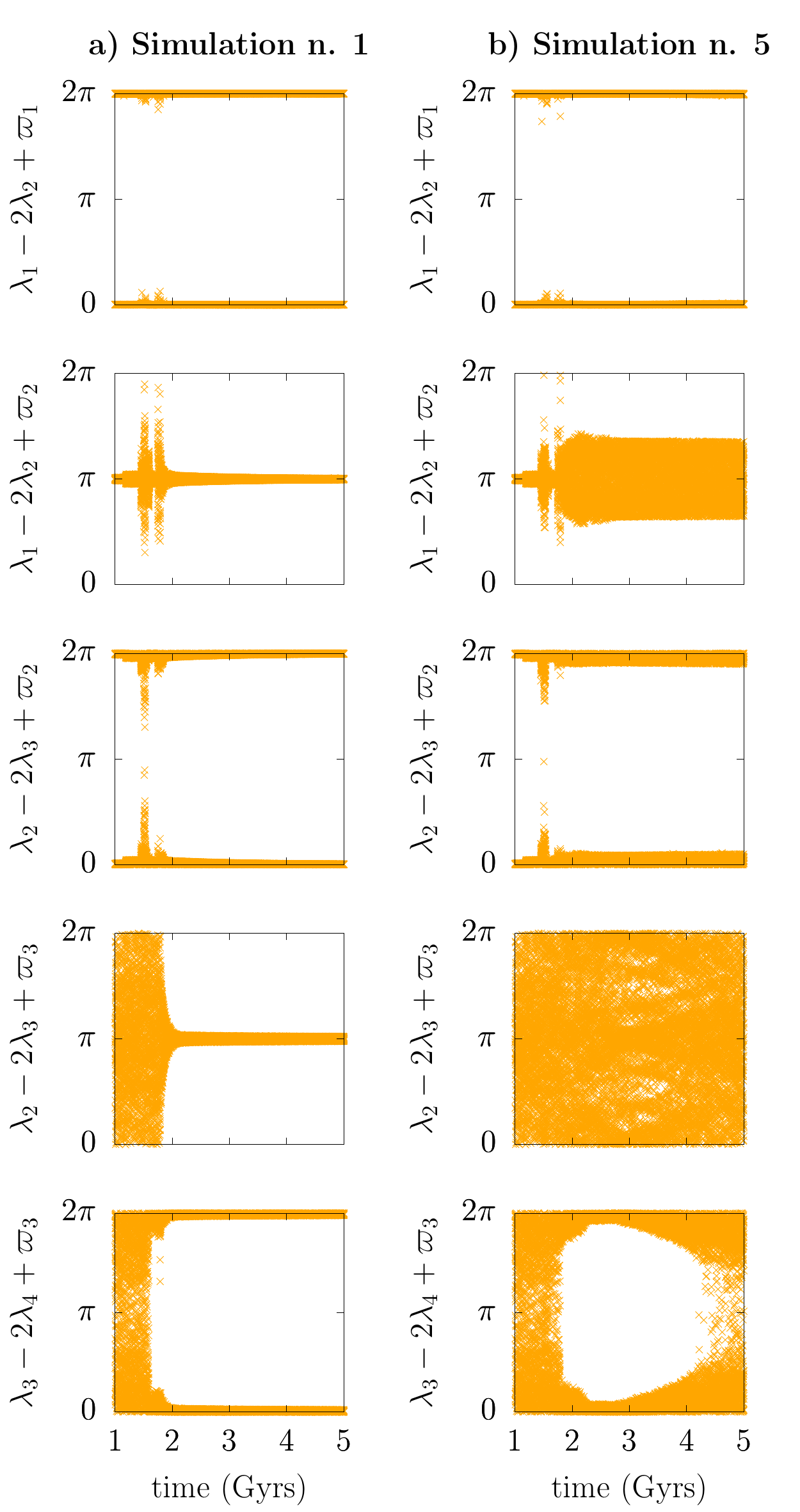}
   \caption{Typical evolution of the first-order resonant angles in case A. Column (a): $\lambda_2-2\lambda_3+\varpi_3$ starts to librate. Column (b): $\lambda_2-2\lambda_3+\varpi_3$ continues to circulate. See text for the definition of case A and a description of the dynamics.}
   \label{fig:angcasea}
\end{figure}

\begin{figure}
   \centering
   \includegraphics[scale=0.72]{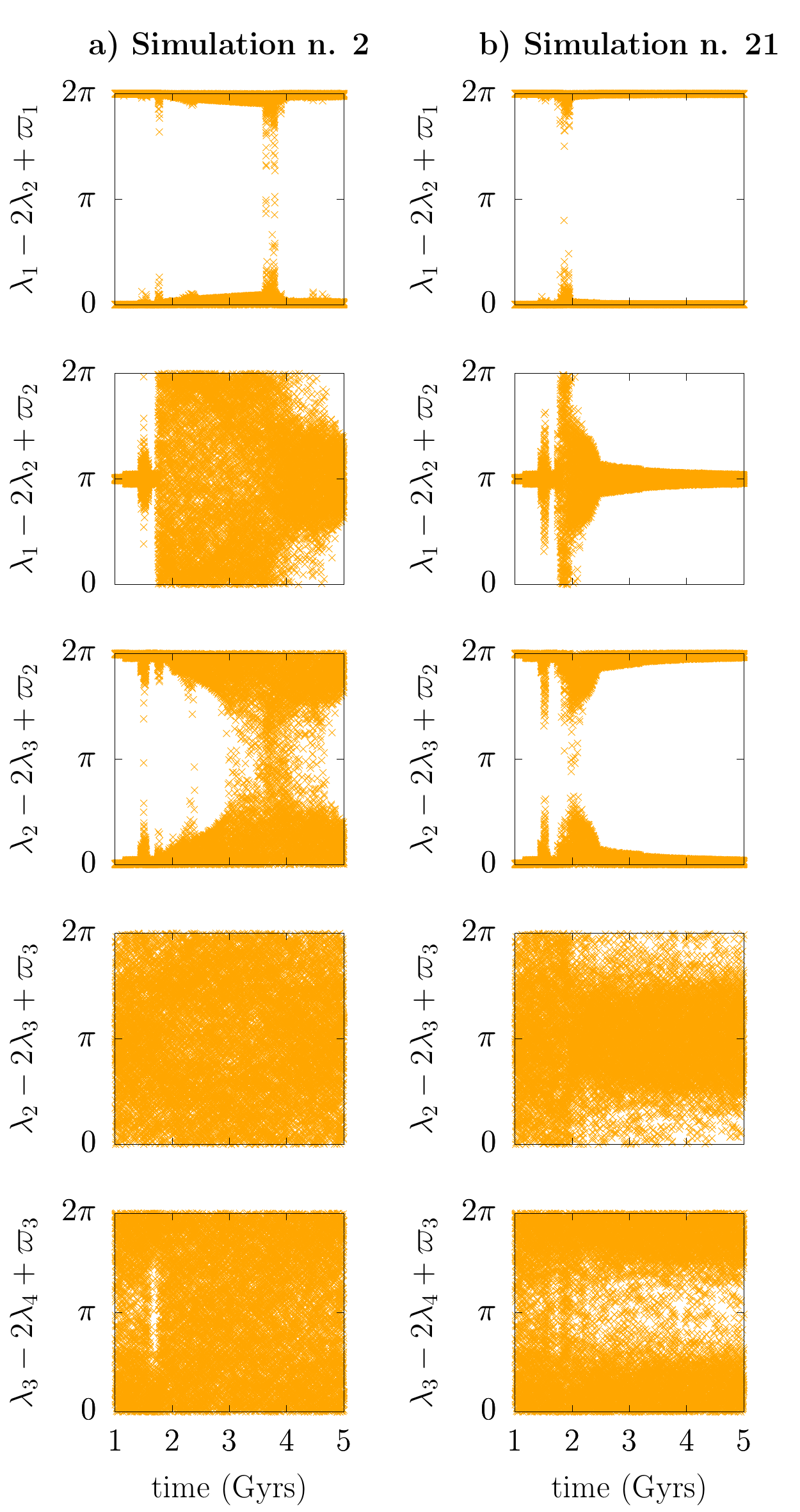}
   \caption{Typical evolution of the first-order resonant angles in case B. Column (a): $\lambda_2-2\lambda_3+\varpi_2$ starts to circulate. Column (b): $\lambda_2-2\lambda_3+\varpi_2$ continues to librate. See text for the definition of case B and a description of the dynamics.}
   \label{fig:angcaseb}
\end{figure}

As a general result of our $628$ simulations, Callisto always ends up captured in some mean-motion resonance. Indeed, a secular drift of its semi-major axis is triggered in all simulations, implying that the dissipative effects on Io manage to reach the orbit of Callisto through some chain of mean-motion resonances. However, at about $1.5$~Gyr,  our simulations split into different cases. We classify them according to the end state of the system. We discriminate between
\begin{itemize}
   \item case A: a chain of three 2:1 two-body mean-motion resonances in the couples Io--Europa, Europa--Ganymede, and Ganymede--Callisto; and
   \item case B: a resonant chain including the 2:1 mean-motion resonance between Io and Europa, plus at least one pure three-body resonance.
\end{itemize}
As in \cite{GALLARDO-etal_2016}, a ``pure'' three-body resonance means that it is not the result of the sum of two-body resonances (contrary to the current configuration of Io, Europa and Ganymede). Therefore, it corresponds to the librations of a three-body resonance angle while the corresponding two-body angles circulate. For instance, if $\sigma = \sigma_1+\sigma_2$ is a librating three-body-resonance angle, it is referred to as pure if the two-body-resonance angles $\sigma_1$ and $\sigma_2$ both circulate (two examples are given in \figurename~\ref{fig:ang3b}). By observing the drift of semi-major axis, we can be assured that the pure three-body mean-motion resonance indeed drives the dynamics.

Cases A and B cover our whole set of $628$ simulations. Within them, we can distinguish different behaviors by looking at the evolution of the resonant angles and eccentricities. The Laplace resonance that remains stable up to about $1.5$~Gyr can be preserved or disrupted, as we see below. However, it is worth noting that the only angle that continues to librate in all simulations is $\lambda_1-2\lambda_2+\varpi_1$ (although it can be temporarily excited as a result of the creation or disruption of other resonances). This means that the couple Io--Europa always remains locked in the 2:1 resonance. Indeed, Io and Europa are further away from Callisto than Ganymede and their dynamics are less perturbed by the resonant encounter. We also note that the inclination degrees of freedom appear to be unimportant in this problem: the inclinations remain low, even though we did not include any damping of their values (see Sect.~\ref{subsec:tides}), and no major inclination resonance is found to drive the dynamics in any of our $628$ simulations.

\begin{figure*}[t]
   \centering
   \includegraphics[scale=0.7]{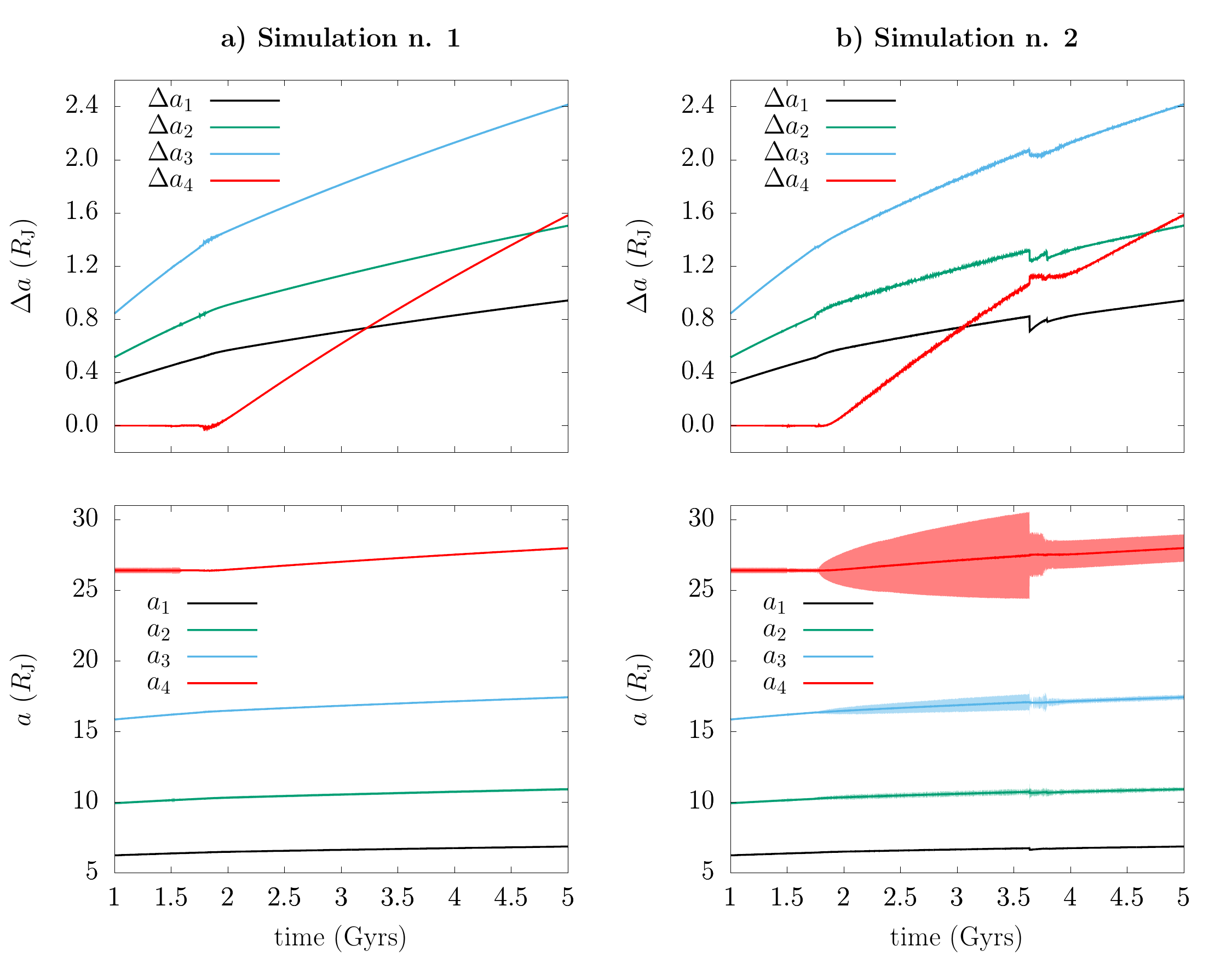}
   \caption{Typical long-term evolution of the semi-major axes ($\Delta a$ and $a$). The bottom graphs also show the pericenter and apocenter distances, represented as a colored interval around the value of $a$. The left column shows a stable case where, after the first capture of Callisto into resonance, the system remains in the same configuration and the migration of the satellites is almost linear. The right column shows an unstable case where, at about $3.5$~Gyr after time J2000, one of the resonances is disrupted and a new one is formed.}
   \label{fig:axes_evol}
\end{figure*}

\subsection{Case A: two-body resonant chain}\label{subsec:caseA}
Case A is the most probable outcome ($354$ simulations over $628$): Ganymede and Callisto enter into a 2:1 two-body mean-motion resonance, while the current resonances between Io, Europa, and Ganymede are preserved (see Eq.~\ref{lap2bod}), as well as the Laplace relation (see Eq.~\ref{lapang}). The mean longitudes of Ganymede and Callisto verify
\begin{equation}
   \label{resgaca}
   \lambda_3-2\lambda_4+\varpi_3 \sim 0,
\end{equation}
as shown in Fig.~\ref{fig:angcasea}. The new resonant angle is of the first order in the masses and in the eccentricities, therefore it is a very strong term in the Hamiltonian. The angle $\lambda_3-2\lambda_4+\varpi_4$ also happens to librate in some simulations ($72$ over $354$), but never without Eq.~\eqref{resgaca}, and this does not affect the qualitative behavior of the system.

The resonance between Ganymede and Callisto completes the full chain of 2:1 resonances, such that once Callisto is captured, it starts to migrate outward (see \figurename~\ref{fig:axes_evol}a). This shows that the dissipative effects acting on the orbit of Io spread to all moons and now reach Callisto. Figure~\ref{fig:ecc_evol}a shows that after the crossing of the chaotic region generated by the resonant encounter, the eccentricities stabilize to new low values forced by the two-body resonances. These values remain below $0.01$, similar to the ones we observe nowadays, along the whole propagation of $5$~Gyr.

In most simulations ending in case A ($326$ over $354$), another angle begins to librate:
\begin{equation}
   \label{desitter}
   \lambda_2-2\lambda_3+\varpi_3 \sim 0\,,
\end{equation}
as illustrated in Fig.~\ref{fig:angcasea}a. This is the missing relation that defines the De Sitter resonance, allowing the existence of periodic orbits for the four-body system composed of Jupiter, Io, Europa, and Ganymede (see \citealp{DESITTER_1909}). This additional resonance means that the longitudes of the satellites' nodes all precess at the same rate: we have $\varpi_2-\varpi_1\sim \pi$, and $\varpi_3-\varpi_2\sim \pi$. This also implies that five of the six first-order resonance angles librate (we have simultaneously Eqs.~\ref{lap2bod}, \ref{resgaca}, and \ref{desitter}). This is a very stable configuration: once the eccentricities are settled in their new forced values, our integrations do not show any significant change. The satellites continue to migrate outward and all the established resonances are preserved. The simultaneous Eqs.~\eqref{lap2bod}, \eqref{resgaca}, and \eqref{desitter} imply that a large number of other angles librate, including
\begin{equation}
   \label{4bod}
   \lambda_1-2\lambda_2-\lambda_3+2\lambda_4\sim 0\,.
\end{equation}
Like the current Laplace resonance (see Eq.~\ref{lapang}), this last relation is a geometrical consequence of the libration of other angles. It means that when Io and Ganymede are in conjunction, so must be Europa and Callisto, a very interesting configuration that involves all the Galilean satellites.

In a few simulations ending in case A ($28$ over $354$), on the contrary, the angle $\lambda_2-2\lambda_3+\varpi_3$ continues to circulate (compare \figurename~\ref{fig:angcasea}a and b). In this case, we observe that the 2:1 resonance between Ganymede and Callisto can be disrupted after a few billion years (i.e., Eq.~\ref{resgaca} is no longer verified). Indeed, $\lambda_3-2\lambda_4+\varpi_3$ oscillates with a wider and wider amplitude until it returns to circulation, and Europa, Ganymede, and Callisto eventually end up in a pure three-body resonance. This evolution is characterized by a slow increase of Callisto's eccentricity (see \figurename~\ref{fig:ecc_evol}b), which stops once the system settles in its new configuration. However, this process is extremely slow.

\subsection{Case B: Chain with a pure three-body resonance}\label{subsec:caseB}

The remaining simulations ($274$ over $628$) show more complex evolutions involving the formation of a 4:2:1 pure three-body mean-motion resonance. Theoretically, this kind of resonance could involve the triplet Io--Europa--Ganymede, or the triplet Europa--Ganymede--Callisto, or both of them. However, the pure resonance Io--Europa--Ganymede only appears as a transitory state in our simulations (see Sect.~\ref{subsec:lapevol} below). We only found one simulation in which a pure resonance Io--Europa--Ganymede seemed to have lasting effects, but due to its low statistical significance, and since the evolution of the eccentricities in this simulation does not differ much from the general case B described below, we do not emphasize it any further. All the other simulations classified in case B ($273$ over $274$) involve a pure three-body resonance between Europa, Ganymede, and Callisto. Differently from case A, Ganymede and Callisto do not lock into the 2:1 two-body resonance (see Fig.~\ref{fig:angcaseb}), at least not immediately, but they enter into a pure three-body resonance with Europa. As before, all simulations show a drift in the  semi-major axis of Callisto, which asserts its capture into resonance.

\begin{figure*}[t]
   \centering
   \includegraphics[scale=0.98]{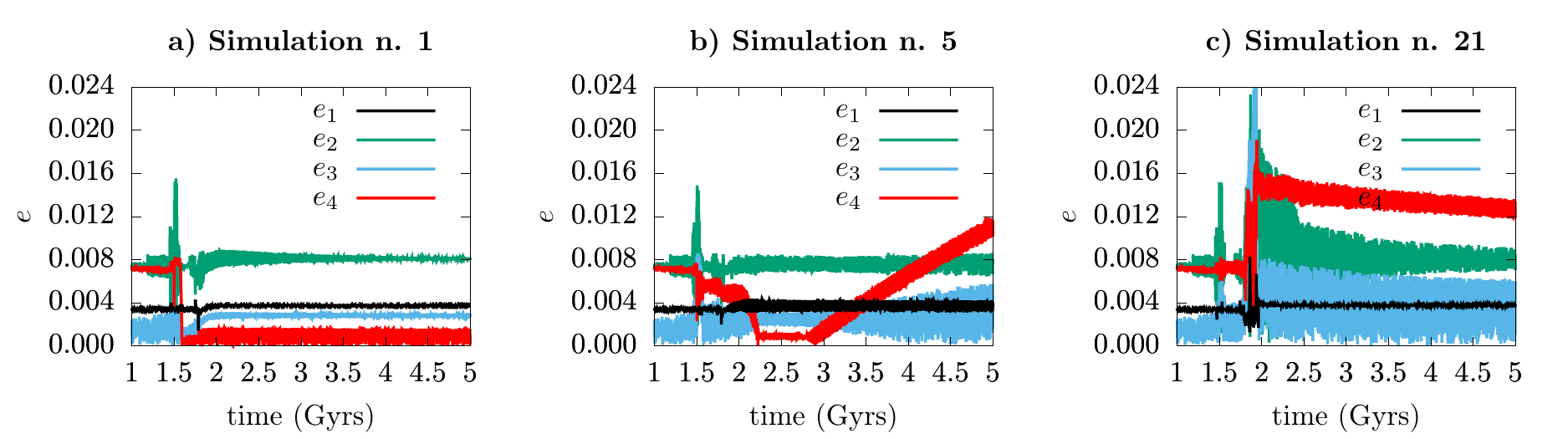}
   \caption{Typical evolution of the eccentricities in simulations where the Laplace resonance and all the current resonances are preserved. Column (a): Case A with $\lambda_2-2\lambda_3+\varpi_3$ in libration; all the eccentricities rapidly settle to new low values. Column (b): Case A with $\lambda_2-2\lambda_3+\varpi_3$ in circulation; Callisto's eccentricity increases slowly as it exits its resonance with Ganymede. Column (c): Case B with $\lambda_2-2\lambda_3+\varpi_2$ in libration; after an abrupt increase the eccentricities converge to new values.}
   \label{fig:ecc_evol}
\end{figure*}

\begin{figure}
   \centering
   \includegraphics[scale=0.7]{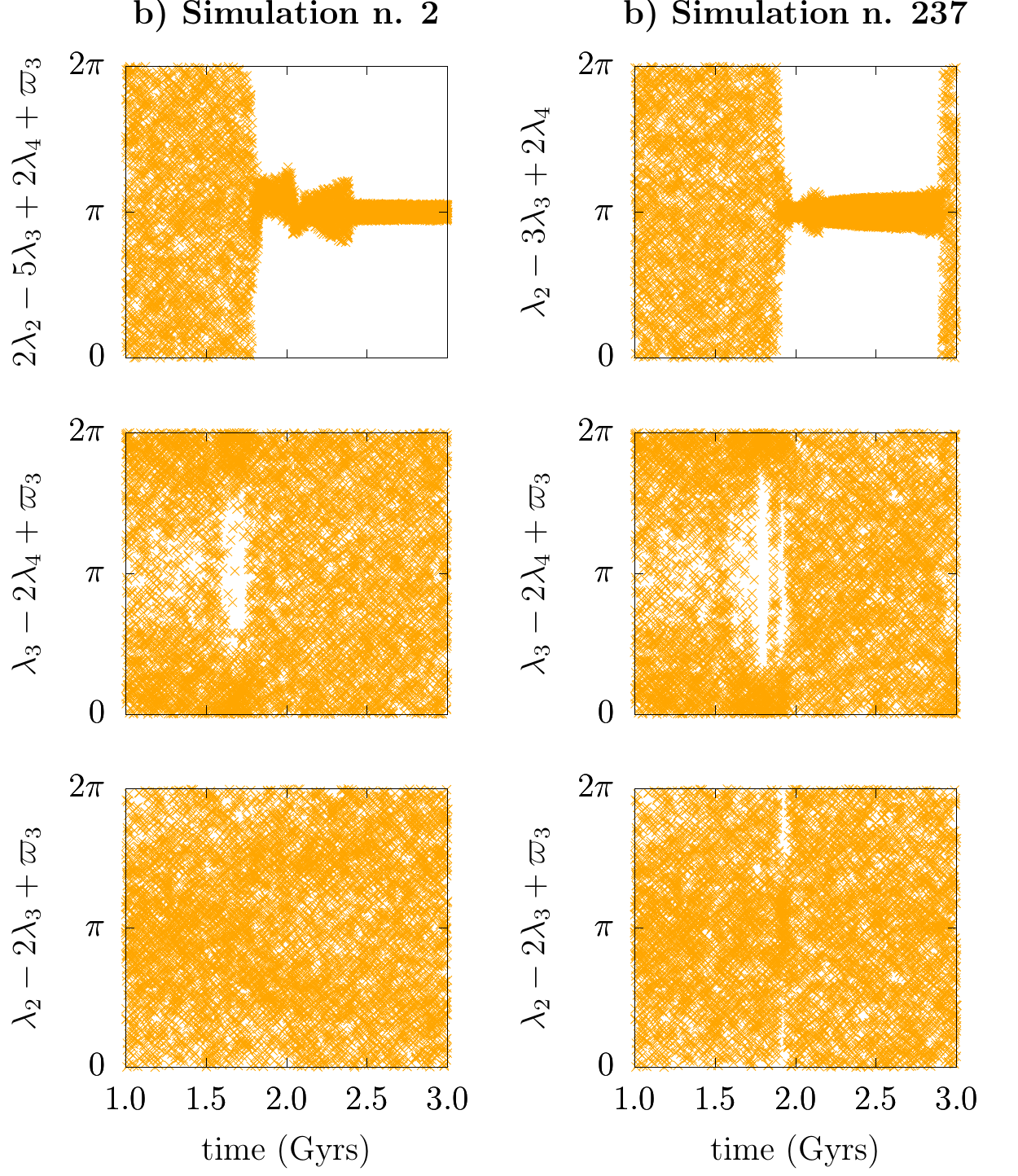}
   \caption{Examples of simulations in which Callisto is trapped in a pure three-body resonance: on the left, a case with $2\lambda_2-5\lambda_3+2\lambda_4+\varpi_3\sim\pi$; on the right, one with $\lambda_2-3\lambda_3+2\lambda_4\sim\pi$.}
   \label{fig:ang3b}
\end{figure}

Most simulations classified in case B ($212$ over $274$) are characterized by the resonant angle
\begin{equation}\label{eq:res3b1}
   2\lambda_2-5\lambda_3+2\lambda_4+\varpi_3\sim \pi \,,
\end{equation}
and a few others ($48$ over $274$) have
\begin{equation}\label{eq:res3b2}
   \lambda_2-3\lambda_3+2\lambda_4\sim \pi \,.
\end{equation}
Typical examples are given in Fig.~\ref{fig:ang3b}. The remaining simulations classified in case B involve other three-body resonances that are not always easy to identify. The terms associated to these angles are of the second order in the masses. This means that they do not directly appear inside the Hamiltonian in Eq.~\eqref{eq:H}; instead, they appear in the remainders of the Lie-series when using a perturbative approach (see e.g., \citealp{NESVORNY-MORBIDELLI_1998}). By computing these remainders, we observe that the lowest-order three-body angles result from the sum or the difference of two circulating two-body angles (see Eq.~\ref{eq:HMres}). In our simplified model, such terms are at least of order two in the eccentricities. These terms are relatively small, but they are incredibly numerous; and indeed, we observe that all pure three-body resonances in our model appear when $\varpi_3-\varpi_2$ and/or $\varpi_4-\varpi_3$ librate, that is, when numerous combinations analogous to Eqs.~\eqref{eq:res3b1} and \eqref{eq:res3b2} act together and combine their effects. This property is discussed further below.

At this point, it is worth noting that in the process of eliminating short-period terms from the Hamiltonian (see Sect.~\ref{sec:model}), we eliminated many three-body resonant combinations. For example, the fast angles $\lambda_2-\lambda_3$ and $2\lambda_3-2\lambda_4$ that are absent from our model would generate a contribution to Eq.~\eqref{eq:res3b2} of order zero in eccentricity. More generally, a complete nonaveraged dynamical model would contain more pure three-body resonances than our model. On the one hand this would increase the capture probability of Callisto (which is already $100\%$ in our simulations), but on the other hand it could  somehow alter the classification scheme that we use, especially concerning the simulations ending in case B. Therefore, the simulations presented below are not meant to be representative of every possible evolution involving pure three-body resonances. However, we highlight the fact that our results feature the same three-body inequalities as those obtained by \cite{MALHOTRA_1991} and \cite{SHOWMAN-MALHOTRA_1997}. For instance, the Laplace-like resonance of these latter authors, identified by $(2n_2-n_1)/(2n_3-n_2)\approx 1/2,$ can be rewritten as $2n_1-5n_2+2n_3\approx 0$, which is the same relation obtained here for the three outer satellites by deriving Eq.~\eqref{eq:res3b1}. The evolution of the satellites' eccentricities, which is described below, is also very similar.

\begin{figure*}
   \centering
   \includegraphics[scale=0.98]{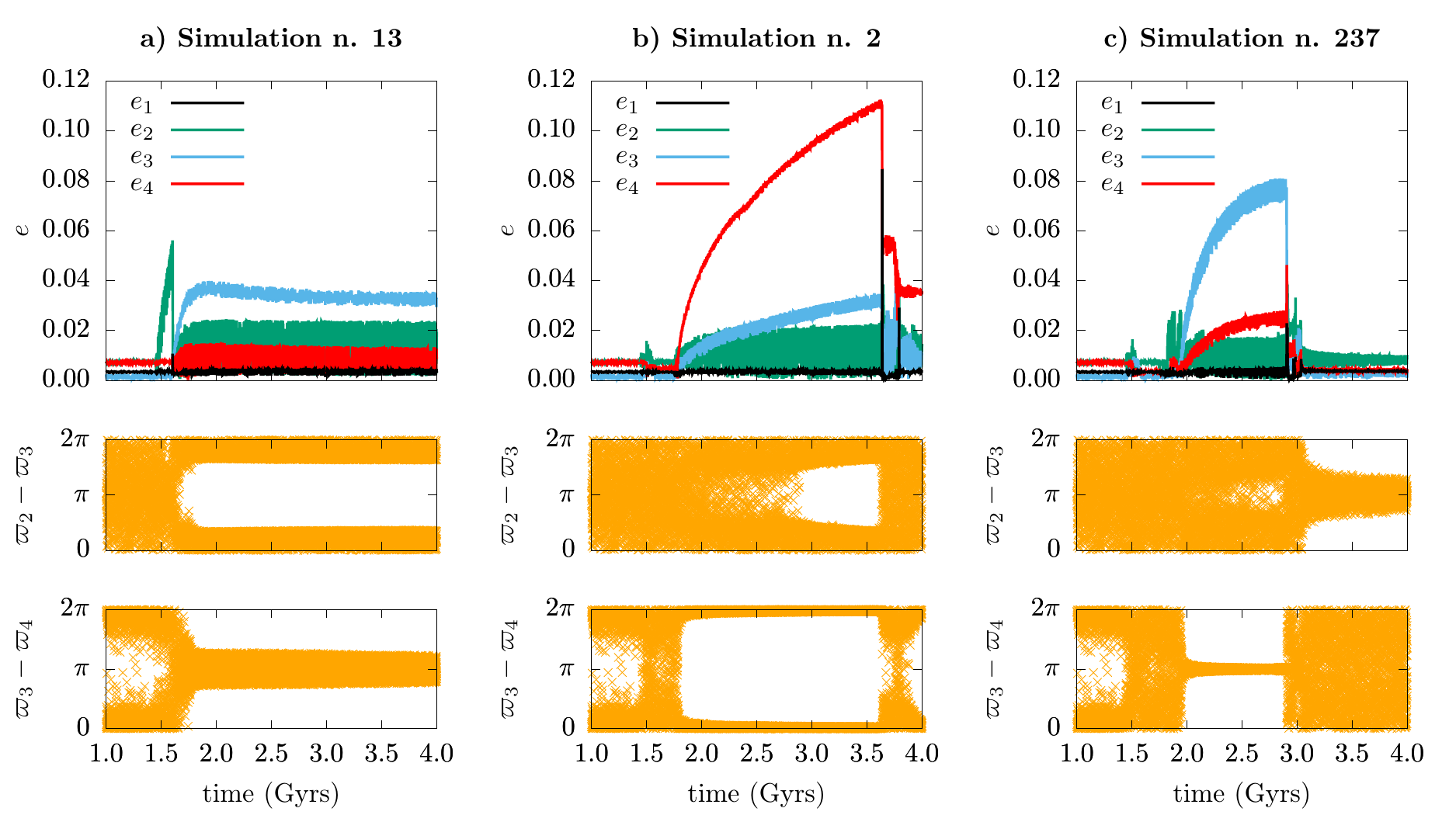}
   \caption{Typical evolution of the eccentricities in simulations where the Laplace resonance is disrupted. Column (a): Case B with $\varpi_2-\varpi_3\sim 0$; all the eccentricities remain below $0.04$. Column (b): Case B with $\varpi_2-\varpi_3$ in circulation and $\varpi_3-\varpi_4\sim 0$; the eccentricity of Callisto increases until the pure three-body resonance is disrupted. Column (c): Case B with $\varpi_2-\varpi_3$ in circulation and $\varpi_3-\varpi_4\sim \pi$; the eccentricity of Ganymede increases until the pure three-body resonance is disrupted.}
   \label{fig:ecc_evol2}
\end{figure*}

In most simulations classified in case B ($233$ over $274$), the resonances $\lambda_2-2\lambda_3+\varpi_2$ and $\lambda_1-2\lambda_2+\varpi_2$ are destroyed. The 2:1 resonance between Io and Europa is the only resonance that survives (see Eqs.~\ref{lap2bod} and \ref{lapang}), while the pure three-body resonance appears. The Laplace resonance is broken, destabilized by the resonant encounter with Callisto. This transition can be slow (about $1$~Gyr, as shown in \figurename~\ref{fig:angcaseb}a) or very fast (a few million years). During this transition, the eccentricities of the satellites evolve in strong correlation with the longitudes of their pericenters:
\begin{itemize}
   \item If $\varpi_2-\varpi_3$ librates, the eccentricity of Ganymede increases quickly up to about $0.04$. Then the whole system stabilizes, and the three-body resonance is preserved up to the end of the five-gigayear integration (see \figurename~\ref{fig:ecc_evol2}a).
   \item If $\varpi_2-\varpi_3$ circulates, but $\varpi_3-\varpi_4$ librates, the eccentricities of Ganymede and Callisto slowly increase up to large values. A similar evolution was observed by \cite{MALHOTRA_1991} and \cite{SHOWMAN-MALHOTRA_1997} before the formation of the current Laplace resonance. We observe distinct behaviors of the eccentricities according to the value around which $\varpi_3-\varpi_4$ librates (see \figurename~\ref{fig:ecc_evol2}b and c). Its libration around zero produces a faster increase of the eccentricity of Callisto, while its libration around $\pi$ produces a faster increase of that of Ganymede. This happens because the three-body resonant terms that dominate are not the same in both cases. This is similar to the mechanism described by \cite{PICHIERRI-etal_2019}: as energy is gradually dissipated, the satellites adiabatically follow the resonance center, which drifts to higher and higher values of the eccentricities. However, beyond some threshold of the eccentricities, the system appears to be unstable. This is probably because the increase of the eccentricities widens neighbor resonances, which eventually overlap and destabilize the system. The pure three-body resonance is therefore disrupted and eccentricities are damped again to very small values. The satellites are then immediately captured into a new resonant configuration, which cannot be uniquely determined because of the chaotic
nature of the transition. As shown in Fig.~\ref{fig:ecc_evol2}, these cycles can go on for billions of years.
\end{itemize}

In the remaining simulations classified in case B ($40$ over $274$), $\lambda_2-2\lambda_3+\varpi_2$ continues to librate (see \figurename~\ref{fig:angcaseb}b). Therefore, Europa and Ganymede remain locked in their two-body resonance and the Laplace relation~\eqref{lapang} remains, while Callisto enters into a pure three-body resonance with Europa and Ganymede. Since the current resonances between Io, Europa, and Ganymede are preserved, the variations of their eccentricities remain moderate, as shown in \figurename~\ref{fig:ecc_evol}c. The eccentricity of Callisto is the only one to suffer from a slight increment, but then it stabilizes rapidly below $0.02$. For some of these simulations, we observe a slow transition to case A: after a few billion years, Ganymede and Callisto finally enter the 2:1 two-body resonance.

Throughout this section, we see that simulations classified as case B can feature a large increase of the eccentricity of Ganymede and/or Callisto (up to about $0.1$). However, as illustrated in Fig.~\ref{fig:axes_evol}b,  these growths are far too small to allow them to cross orbits of other satellites: this prevents any catastrophic event, such as ejections or collisions.

   
\section{Discussion}\label{sec:discuss}

\subsection{Evolution of the Laplace resonance}\label{subsec:lapevol}
Section~\ref{sec:evol} shows that the resonant encounter with Callisto can preserve the Laplace resonance between Io, Europa, and Ganymede (case A and a few simulations from case B), or destroy it (case B).  More precisely, the Laplace resonance, meant as the chain between the 2:1 resonances of the couples Io--Europa and Europa--Ganymede, is preserved in $394$ over $628$ simulations (about $63\%$). In the remaining simulations, this configuration is destroyed: the angles $\lambda_1-2\lambda_2+\varpi_2$ and $\lambda_2-2\lambda_3+\varpi_2$ pass from libration to circulation, and the relation in Eq.~\eqref{lapang} no longer holds.

\begin{figure}[t]
   \centering
   \includegraphics[scale=0.54]{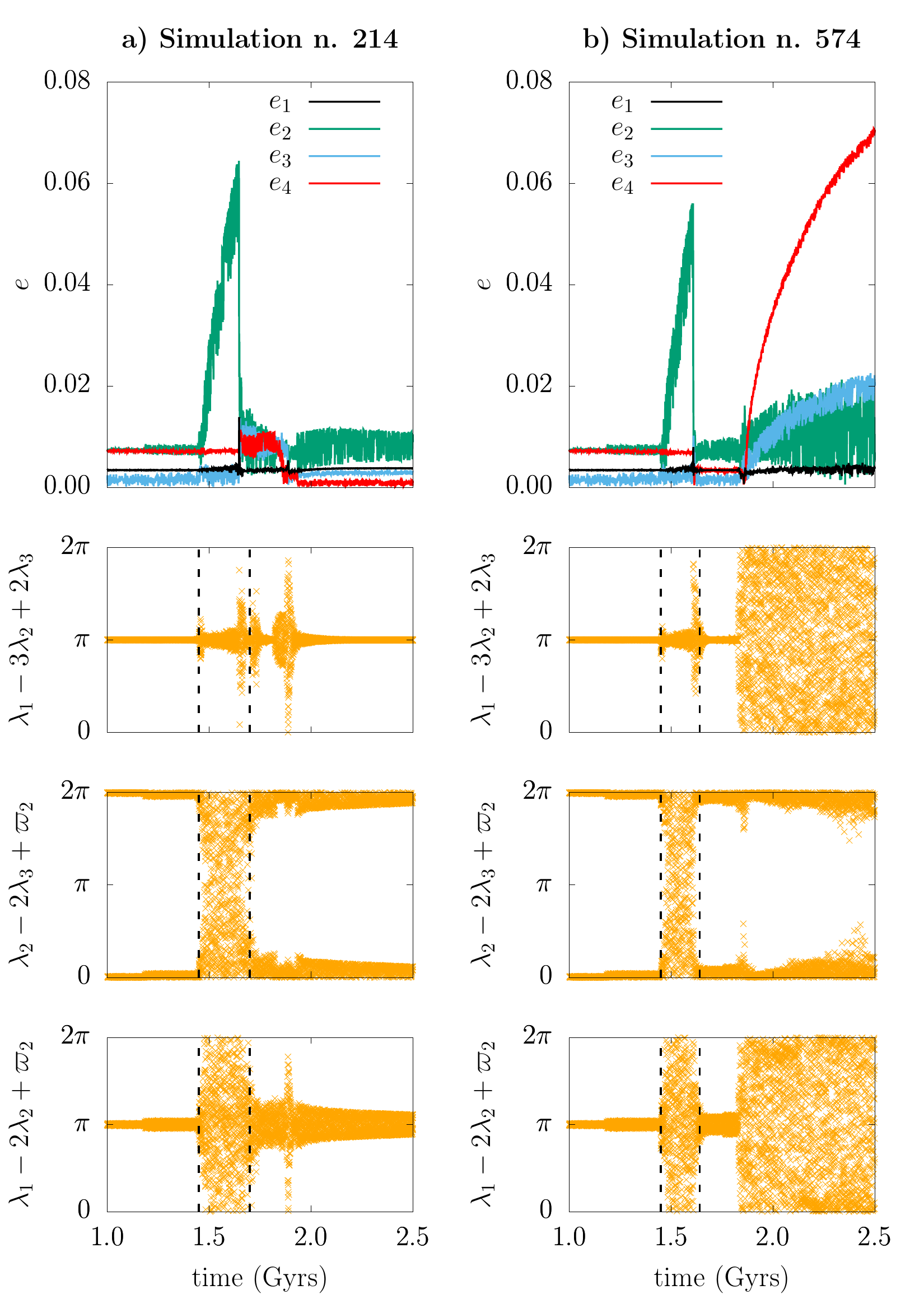}
   \caption{Examples of simulations where the three-body resonance between Io, Europa, and Ganymede becomes pure for a few hundred million years. The area confined between the two dashed black lines is the time span where $\lambda_1-2\lambda_2+\varpi_2$ and $\lambda_2-2\lambda_3+\varpi_2$ circulate, and $\lambda_1-3\lambda_2+2\lambda_3$ librates. Left: Transition to case A. Right: Transition to case B.}
   \label{fig:caselap}
\end{figure}

Nonetheless, for a restricted period of time during the chaotic transitions observed in cases A and B, we found a few examples in which the two-body angles $\lambda_1-2\lambda_2+\varpi_2$ and $\lambda_2-2\lambda_3+\varpi_2$ start to circulate while the three-body relation~\eqref{lapang} still holds. This means that the 4:2:1 three-body resonance between Io, Europa, and Ganymede becomes pure. This configuration generally persists for only a few hundred million years. As shown in \figurename~\ref{fig:caselap}, this ``pure Laplace resonance'' induces a peculiar evolution of the eccentricities: that of Europa shows a rapid and significant increment up to $0.06$, while those of the other moons remain anchored to low values. This mechanism is similar to the one described in Sect.~\ref{subsec:caseB} (case B), which makes the eccentricities of Ganymede and Callisto increase when the three outer satellites are locked in a pure three-body resonance. This is also what \cite{MALHOTRA_1991} and \cite{SHOWMAN-MALHOTRA_1997} obtained while studying the formation of the Laplace resonance.

\subsection{The jungle of two- and three-body resonances}\label{subsec:jungle}
\begin{figure}
   \centering
   \includegraphics[scale=0.9]{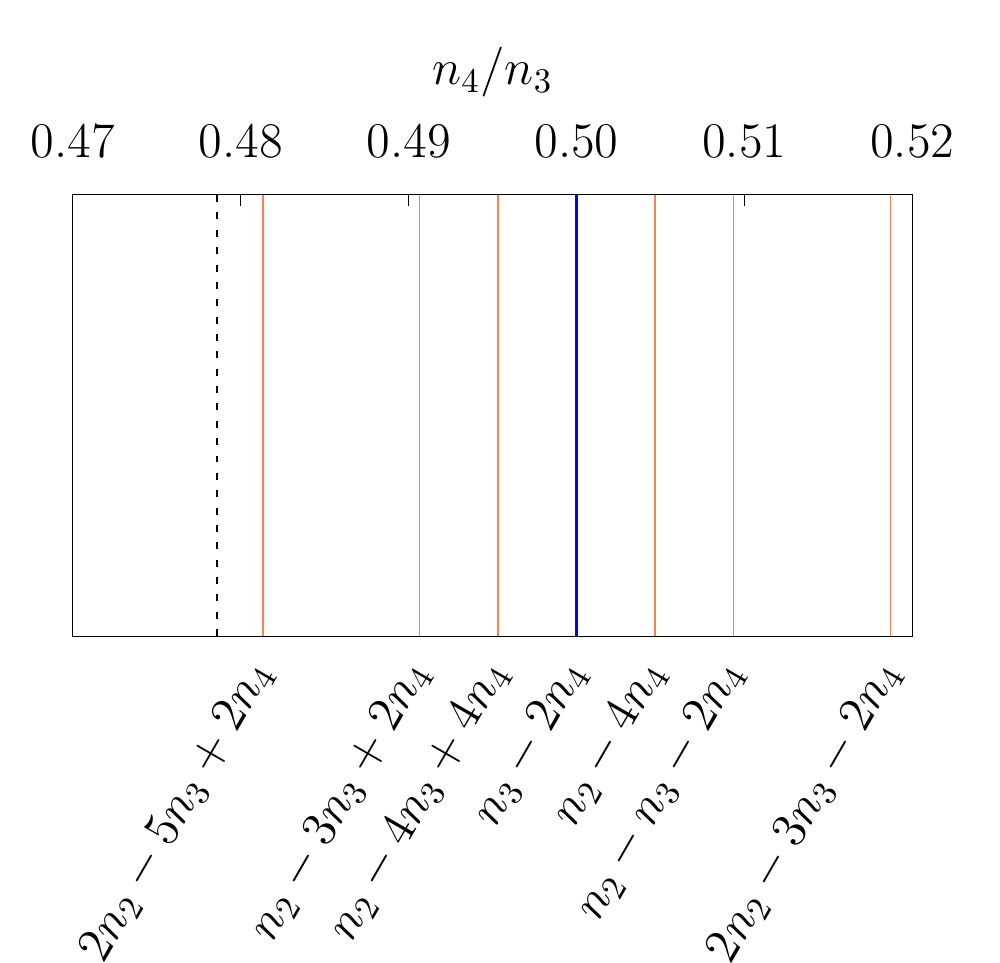}
   \caption{Location of the two-body (blue) and three-body (red) mean-motion resonances as a function of the ratio between the mean motions of Callisto and Ganymede. The dashed black line is its value at $1.4$~Gyr. Tidal dissipation makes it move from left to right.}
   \label{fig:resloc}
\end{figure}

Section~\ref{sec:evol} shows that due to tidal dissipation, numerous two-body and three-body mean-motion resonances can affect the orbital dynamics of the Galilean satellites in the future. Such resonances do not appear randomly. Since they mainly depend on the period ratios among the satellites (and not much on their precession rates), it is even possible to roughly estimate their location. As Io, Europa, and Ganymede are initially tightly locked in resonance (i.e., their period ratios are fixed), the different resonances can be located as a function of Callisto's period ratio only, for instance with respect to Ganymede. From the Hamiltonian function in Eq.~\eqref{eq:HMres}, the only possible three-body resonances at second order of the masses are of the form
\begin{equation}
   \begin{aligned}
       (n_2-2n_3) &\pm  (n_3-2n_4) \,,\\
      2(n_2-2n_3) &\pm  (n_3-2n_4) \,,\\
   \end{aligned}
   \hspace{0.5cm}
   \begin{aligned}
       (n_2-2n_3) &\pm 2(n_3-2n_4) \,,\\
      2(n_2-2n_3) &\pm 2(n_3-2n_4) \,.\\
   \end{aligned}
\end{equation}
Figure~\ref{fig:resloc} shows the relative locations of these resonances and the order in which they can be encountered as Io, Europa, and Ganymede migrate outwards. When taking into account the precession rates of the orbits, each of these resonances splits into a series of multiplets that partially overlap with each other, producing the chaotic evolution observed in the simulations (see \citealp{NESVORNY-MORBIDELLI_1998,GALLARDO-etal_2016}). This explains why chaos appears before actually reaching the 2:1 two-body resonance between Ganymede and Callisto. However, if $\varpi_3-\varpi_4$ and/or $\varpi_2-\varpi_3$ oscillate with a small amplitude, many multiplets merge together (exact overlap), allowing the three-body resonance to stand on its own and produce the dynamics described in Sect.~\ref{subsec:caseB} (case B). As shown by \figurename~\ref{fig:resloc}, the first three-body resonance reached by the satellites is $2\lambda_2-5\lambda_3+2\lambda_4$; this resonance is the one that we most frequently find in case B. In case A, on the contrary, the chaotic zone is crossed quickly and the satellites end up in the strong two-body mean-motion resonance.

\section{Significance of our results}\label{sec:dtides}
As detailed in Sect.~\ref{sec:model}, our model inevitably relies on many simplifications. In particular, the results described in Sect.~\ref{sec:evol} are obtained using constant dissipation parameters and through a procedure of computational acceleration (see Sect.~\ref{subsec:tides}). Moreover, the statistical picture of the different dynamical outcomes is developed from a limited number of simulations, which necessarily limits its precision. All these factors are linked and impact our results in some way. Our approach can legitimately be questioned, and its range of validity needs to be investigated. This is the purpose of this section.

\subsection{Acceleration factor}\label{subsec:accel}
Even though our results are presented in Sect.~\ref{sec:evol} in terms of the real physical time $t$, they are obtained by applying an acceleration factor $\alpha=10^2$ to the dissipation parameters. Because of the adiabatic nature of the energy dissipation, this amounts to using an integration time-variable $\tilde{t}$ for the numerical computations, which is related to the physical time through $t\approx\alpha\,\tilde{t}$. As stressed in Sect.~\ref{subsec:tides}, this method is relevant only as long as the accelerated dissipation process is still adiabatic. If not, we expect spurious artifacts to appear in the dynamics. The adiabatic nature of the dissipation can be studied by varying the value of $\alpha$. Indeed, the accelerated dissipation process is still adiabatic if: \emph{i)} in the regular portions of the evolution, changing $\alpha$ only changes the timescale; and \emph{ii)} in the chaotic portions of the evolution, changing $\alpha$ does not change the statistics of the outcomes.

\begin{figure*}
   \centering
   \includegraphics[scale=0.7]{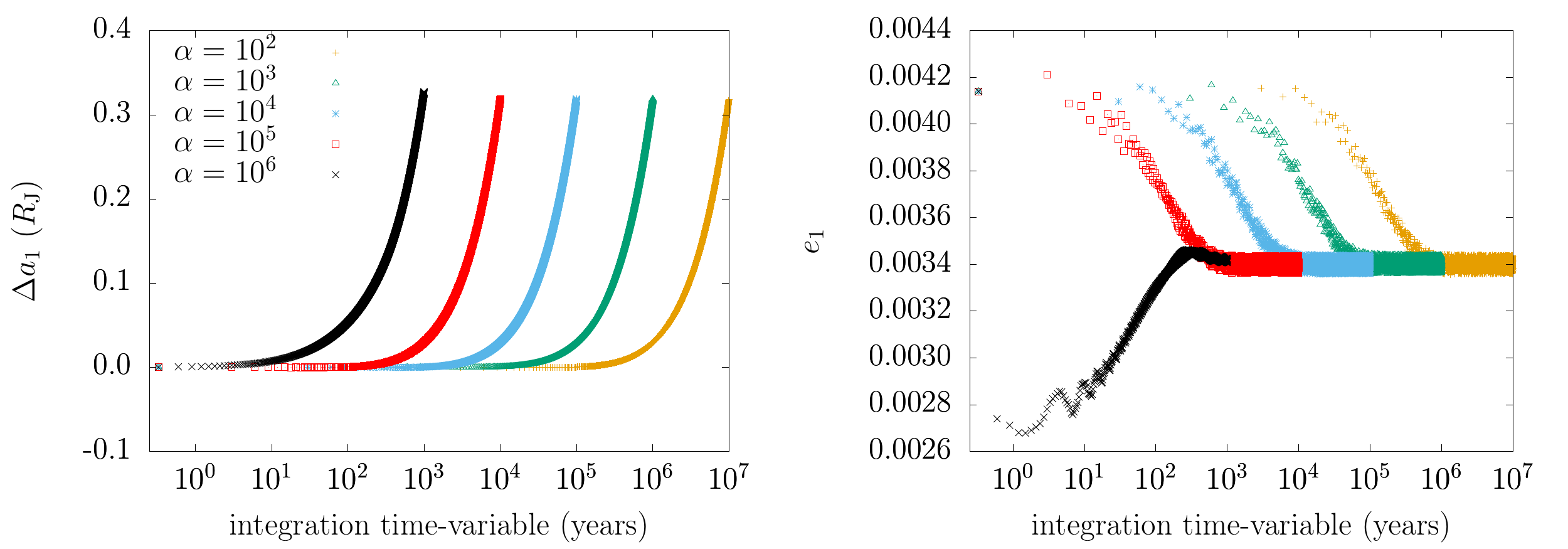}
   \caption{Evolution of the semi-major axis and eccentricity of Io with respect to the integration time-variable $\tilde{t}$ for different acceleration factors $\alpha$. The time is given in a logarithmic scale. The duration of each integration is set so that it represents $1$~Gyr of physical time $t$ used in Sect.~\ref{sec:evol}. Up to $\alpha=10^5$, a change of $\alpha$ only amounts to a linear contraction of the integration time (i.e., the curves overlap when viewed with respect to the physical time $t$).}
   \label{fig:testalpha}
\end{figure*}

For values spanning many orders of magnitude, \figurename~\ref{fig:testalpha} shows the influence of $\alpha$ during the first gigayear of the satellites' evolution (regular dynamics). We only need to examine the semi-major axis and eccentricity of  Io since the energy dissipation is spread through them to the whole satellite system (see Sect.~\ref{subsec:tides}). Figure~\ref{fig:testalpha} shows that no qualitative or quantitative change of the dynamics occurs for values of $\alpha$ ranging up to $10^5$: we only observe a linear contraction of the integration time-variable $\tilde{t}$. For $\alpha=10^6$, on the contrary, the dynamical evolution is completely different: the eccentricity of Io undergoes an abrupt decrease followed by spurious oscillations. Indeed, for $\alpha=10^6$ and beyond, the evolution of $e_1$ is more affected by the magnified dissipation than by the conservative dynamics, meaning that the adiabatic approximation  fails spectacularly.

As a result, the choice of $\alpha=10^2$ seems to be quite reliable, at least during the first portion of the evolution presented in Sect.~\ref{sec:evol}, when the dynamics are regular and driven by the strong Laplace resonance. However, the chaotic transitions that follow feature very weak resonances such as pure three-body resonances. Being shallower, those resonances are associated with longer libration timescales that could endanger the adiabaticity of the accelerated energy dissipation. Although statistical analyses using $\alpha=1$ or $10$ are prohibitive due to overly large computation times, a full statistical picture of the dynamical outcomes can be obtained for larger accelerations. Meaningful statistical deviations beyond a given threshold of $\alpha$ mean that the limit of validity of the adiabatic approximation is reached. This approach has been used for instance by \cite{TITTEMORE-WISDOM_1988}. In order to determine this threshold, we perform $628$ additional simulations for each new value of $\alpha=10^3$, $10^4$, and $10^5$. We then classify them according to the outcome of the resonant encounter with Callisto, as we did in Sect.~\ref{sec:evol}. For better comparison, we enrich our classification scheme as follows:
\begin{itemize}
   \item[A:] Chain of three two-body resonances.
   \item[B.1:] Pure three-body resonance involving Europa, Ganymede, and Callisto.
   \item[B.2:] Pure three-body resonance involving Io, Europa, and Ganymede.
   \item[C:] The two-body resonance between Io and Europa is destroyed.
\end{itemize}
Our results are presented in \tablename~\ref{tab:alpha}. In order to compare them, we first need to quantify the statistical significance of their differences. Assuming that the division between outcomes amounts to a random process, the probability of obtaining $k$ times a given outcome (e.g., case A) among $n=628$ trials obeys a binomial distribution. Using $p$ to denote the probability of obtaining case A when performing a single numerical integration, the expected number of successes is
\begin{equation}
   E = np \,,
\end{equation}
with a variance equal to
\begin{equation}
   \sigma^2 = np(p-1)\,.
\end{equation}
When $n$ grows, the binomial distribution rapidly tends to a normal distribution, so that $E$ and $\sigma^2$ can be interpreted in the usual way. For $n=628$ and a probability $p$ close to $0.5$, the $3\sigma$ uncertainty range of the fraction of successes is about $6\%$. The fractions of cases A and B obtained for $\alpha=10^2$ and $\alpha=10^3$ are therefore perfectly compatible (see Table~\ref{tab:alpha}). For less probable outcomes, the $3\sigma$ range is smaller: for instance, we obtain an uncertainty range of about $3\%$ for $p=0.05$. The fraction of case B.2 obtained for $\alpha=10^2$ and $\alpha=10^3$ are therefore only marginally compatible. This indicates that for an acceleration factor $\alpha=10^3$, the adiabatic approximation already slightly begins to crumple. Finally, the fractions obtained for $\alpha=10^4$ and beyond are clearly not compatible with those obtained for smaller values of $\alpha$; this means that the adiabatic approximation is not valid for such large energy dissipations. In particular, the occurrence of Case C means that Io is pushed so heavily by the dissipation that even the small perturbation due to the resonant crossing of Ganymede with Callisto is able to make it escape its resonance with Europa. We also note that the number of simulations featuring a pure three-body resonance between Europa, Ganymede, and Callisto (Case B.1) abruptly decreases beyond $\alpha=10^4$ because the system has no time to explore such weak resonances before reaching the strong two-body resonance between Ganymede and Callisto (see Sect.~\ref{subsec:jungle} and Fig.~\ref{fig:resloc}). In contrast, the value $\alpha=10^2$ used throughout this article appears to be quite satisfactory.

\begin{table}
\caption{Distribution of outcomes for different acceleration factors $\alpha$.}
\label{tab:alpha}
\centering
\begin{tabular}{r r r r r}
\hline\hline
$\alpha$ & case A & case B.1 & case B.2 & case C \\
\hline
    $10^2$ &  $56.4\%$ &  $43.4\%$ &    $0.2\%$  &    $0.0\%$ \\
    $10^3$ &  $48.4\%$ &  $45.9\%$ &    $5.7\%$  &    $0.0\%$ \\
    $10^4$ &  $56.2\%$ &   $9.2\%$ &   $27.9\%$  &    $6.7\%$ \\
    $10^5$ &  $17.2\%$ &   $0.0\%$ &   $46.0\%$  &   $36.8\%$ \\
\hline                                     
\end{tabular}
\tablefoot{Case A indicates a complete chain of two-body resonances. Case B.1 indicates a pure three-body resonance Europa--Ganymede--Callisto. Case B.2 indicates a pure three-body resonance Io--Europa--Ganymede. Case C indicates the destruction of the two-body resonance Io--Europa.}
\end{table}

\subsection{Tidal dissipation model}\label{subsec:dissmod}
As explained in Sect.~\ref{subsec:tides}, the use of constant dissipation parameters is justified by the fact that their hypothetical variations given by conventional frequency-dependent models remain below the level of uncertainty of their value. In order to explore different dissipation models, it appears therefore more sensible to continue using constant parameters and to sample their respective uncertainty ranges. Due to the adiabatic nature of the dissipation, allowing the parameters to vary would simply mean that the evolution timescales of $a_1$ and $e_1$ are not constant, but that they slowly contract or expand inside the limits given by our sampling. As shown in Sect.~\ref{subsec:accel}, the current dissipation rate could even be multiplied by $10^3$ in the future without threatening its adiabatic nature. For each set of constant parameters, we need to measure critical properties of the dynamics that can serve as a proxy of their effect. Since these parameters mostly modify the evolution timescale, their effects can be quantified by measuring the epoch of the resonant encounter with Callisto. Guided by Fig.~\ref{fig:resloc}, we arbitrarily define the beginning of the resonant encounter when the period ratio of Ganymede and Callisto exceeds $0.48$.

\begin{figure}
   \centering
   \includegraphics[scale=0.85]{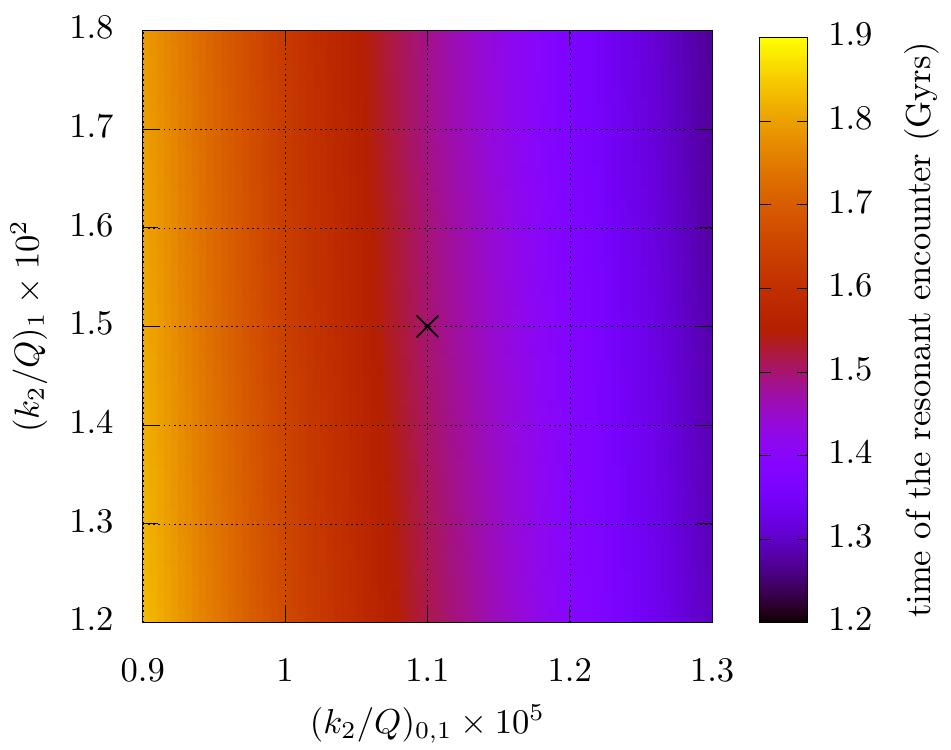}
   \caption{Time of the resonant encounter from today as a function of the values of the dissipative parameters. The axis ranges correspond to the uncertainties given by \cite{LAINEY-etal_2009}.}
   \label{fig:diffevol}
\quad\\
   \includegraphics[scale=0.85]{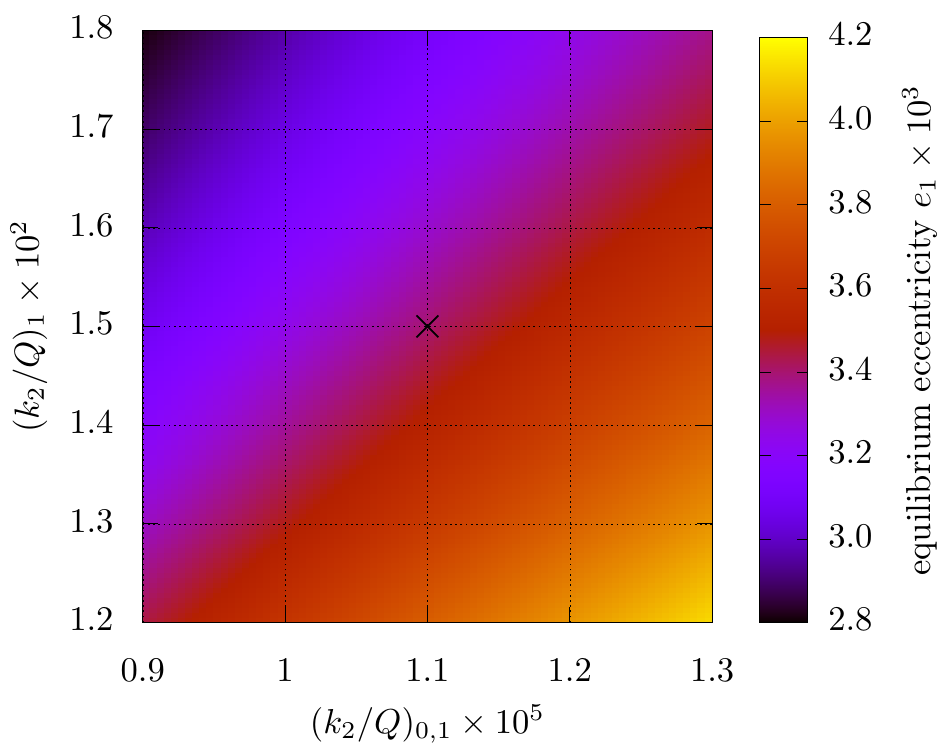}
   \caption{Equilibrium eccentricity of Io before the resonant encounter as a function of the values of the dissipative parameters. The axis ranges correspond to the uncertainties given by \cite{LAINEY-etal_2009}.}
   \label{fig:diffevol_ecc}
\end{figure}

Figure~\ref{fig:diffevol} shows the time of the resonant encounter obtained for a fine grid of parameters $(k_2/Q)_{0,1}$ and $(k_2/Q)_{1}$ sampled within their uncertainties (see Eq.~\ref{eq:dissipIo}). The encounter time is $1.5$~Gyr in our nominal simulations analyzed in Sect.~\ref{sec:evol} (central cross), but we see that it can vary from about $1.2$ to $1.9$~Gyr. The encounter time is much more sensitive to the value of $(k_2/Q)_{0,1}$ than to the value of $(k_2/Q)_{1}$. This is because the dissipation inside Jupiter has a dominant role in ruling the drift of the semi-major axes (see Eq.~\ref{dissa}), especially after the eccentricity of Io stabilizes at a lower value (see \figurename~\ref{fig:ewzoom}). As mentioned in Sect.~\ref{sec:evol}, the convergence value of $e_1$ results from an equilibrium between the resonant dynamics and the tidal dissipation. As shown in \figurename~\ref{fig:diffevol_ecc}, this convergence value is slightly affected by the value of the dissipative parameters sampled in their uncertainty range. This also  somehow modifies the equilibrium eccentricity of Europa (see \figurename~\ref{fig:ewzoom}, where both $e_1$ and $e_2$ vary simultaneously), but not enough to produce noticeable dynamical changes during the resonant encounter with Callisto. Interestingly, the eccentricity of Io is already at its equilibrium value today if ever the dissipative parameters have values $(k_2/Q)_{0,1}=1.3\times 10^{-5}$ and $(k_2/Q)_{1}=1.2\times 10^{-2}$. However, this is at the very limit of the uncertainty range provided by \cite{LAINEY-etal_2009}.

\section{Conclusion}\label{sec:concl}

Tidal dissipation causes the orbits of the Galilean satellites to slowly migrate with time. Energy is mostly dissipated by the tidal interactions between Io and Jupiter, but the effects of the dissipation are then redistributed among the satellites through the Laplace resonance. Over billions of years, this produces an outward migration of Io, Europa, and Ganymede. Since it is not currently involved in any mean-motion resonance, Callisto does not yet migrate substantially. However, as Io, Europa, and Ganymede migrate outwards, Callisto is progressively reached by the 2:1 resonance with Ganymede.

In this article, we study the possible outcomes of this resonant encounter. We focus on the probability of capturing Callisto into mean-motion resonance, and on the stability of the current Laplace resonance. To this end, we used the semi-analytical model of \cite{LARI_2018}, which is adjusted to take into account possible resonances between Ganymede and Callisto, and refined to support numerical integrations over a gigayear timescale. We set the duration of our numerical integrations to $5$~Gyr. We assumed constant dissipation parameters, fixed to the values measured by \cite{LAINEY-etal_2009}. These values are still a matter of debate in the literature, but due to the adiabatic nature of the energy drift, a more detailed dissipation model would mostly change the timescale of the resonant encounter, and not its dynamical properties. The extremely accurate data expected from future space missions (JUICE, Europa Clipper), coupled with astrometric data sets, should provide a better understanding of dissipative parameters \citep{DIRKX-etal_2017,LARI-MILANI_2019}.

We find that up to about $1.5$~Gyr from now, the orbit of Callisto remains virtually unchanged and all the current resonances between Io, Europa, and Ganymede are preserved during their migration. However, after $1.5$~Gyr,  the proximity of the 2:1 mean-motion resonance between Ganymede and Callisto produces chaotic effects and a large variety of outcomes become possible. We draw a statistical picture of the dynamics based on a sample of $628$ integrations.

In $56\%$ of the cases, Callisto is captured right away into the 2:1 resonance with Ganymede (case A). The Galilean satellites therefore reach a perfect chain of two-body resonances. In the remaining $44\%$ of the cases, a resonant chain involving all four satellites is also formed, but it includes a pure three-body 4:2:1 mean-motion resonance (case B). Apart from just one simulation, this three-body resonance involves Europa, Ganymede, and Callisto. From a statistical point of view, we expect an absolute uncertainty of a few percent in the division between cases A and B. In all our $628$ simulations, Callisto remains trapped in some mean-motion resonance, which makes it migrate outwards along with the other satellites. Its capture therefore appears to be a highly probable event. This also suggests that regardless of the tidal history of the Galilean satellites, Callisto never crossed the 2:1 resonance with Ganymede in the past, otherwise it would have remained locked. Indeed, a 2:1 resonance crossing of Ganymede and Callisto without capture would require a huge migration rate, which is incompatible with the observations.

In case A, the eccentricities of all satellites settle to small values. As in the current configuration of the system, the 2:1 resonances force the eccentricities to remain small according to the precession rate of the pericenters (see e.g., \citealp{SINCLAIR_1975}). The tidal dissipation does not greatly affect the value of the forced eccentricities, but it produces a linear drift of the semi-major axes of all four satellites, maintaining the chain of 2:1 period ratios.

In case B, the eccentricities of the satellites can reach large values, especially Ganymede and Callisto (up to about $0.1$). Indeed, once trapped in a pure three-body resonance, the tidal dissipation is found to increase the value of the forced eccentricities, and the satellites adiabatically follow the drift of the resonance center. However, in our simulations, this increase never leads to a total destabilization of the system. Before that, the three-body resonance is disrupted by the large values of the eccentricities; freed from their forced values, the eccentricities are rapidly damped again, allowing for a capture into a new resonance. Since pure three-body resonances are very numerous, these cycles can go on for billions of years. Each capture into a new resonance produces a small jump of the semi-major axes, which are attracted towards the new resonance center before resuming their linear drift.

Our study reveals that the resonant encounter with Callisto can destruct any feature of the Laplace resonance as we know it today, except the 2:1 resonance between Io and Europa (which persists in all our simulations). Hence, the Laplace resonance is stable under the action of tidal dissipation, but not under the resonant encounter with Callisto that happens at about $1.5$~Gyr from now. Even though all four satellites invariably end up in a new resonant chain, the 2:1 resonance between Europa and Ganymede is destroyed in $37\%$ of our simulations. The Laplace resonance can then turn into a pure three-body resonance between Io, Europa, and Ganymede; however, this is a rare outcome of our simulations, and it generally lasts less than a few hundred million years. During this interval of time, the eccentricity of Europa increases.

The orbital inclinations of the satellites are not found to play any role in their long-term dynamics: they remain small at all times and are only slightly affected when the satellites enter into or exit from resonances.

Our approach has two main limitations. At first, since the Hamiltonian is truncated at second order in the eccentricities, our model is less accurate when the eccentricities are large, as in some simulations of case B. This could affect the final outcome of a few of our simulations, but not our classification scheme or the percentages given in this conclusion. More importantly, in the process of averaging the Hamiltonian over fast angles, many pure three-body combinations were removed, and in particular the terms of order zero in the eccentricities. Since we observed that the system can be trapped in numerous weak resonances, the long-term evolution given by a nonaveraged model would probably show even more resonant captures, making the escape of Callisto even more improbable. However, the additional three-body resonances could also contribute to the chaos observed in case B and drive more simulations into case A. The percentages obtained in our study should therefore be taken as indicative. Unfortunately, a statistical study over $5$~Gyr using a nonaveraged model would require prohibitive computation times.
   
   
   \begin{acknowledgements}
      This work was funded in part by the Italian Space Agency (ASI).
      The authors would like to thank the anonymous reviewer for his/her comments, which
      significantly improved the manuscript.
      M.~S. thanks Gwena{\"e}l Bou{\'e} and gives entire credit to him 
      concerning the treatment of the non-inertial nature of the reference 
      frame, as well as for the Poincar{\'e}-style change of variables towards 
      the Jovicentric canonical coordinates (Appendix~\ref{asec:ham}).
      M.~F. has been partially supported by the Marie Curie Initial Training Network Stardust-R,
      grant agreement Number 813644 under the H2020 research and innovation program, and acknowledges
      the project MIUR-PRIN 20178CJA2B titled ``New  frontiers of Celestial Mechanics:
      theory and applications''.
   \end{acknowledgements}


   \bibliographystyle{aa}
   \bibliography{satgal2Callisto}

\begin{thebibliography}{74}
\expandafter\ifx\csname natexlab\endcsname\relax\def\natexlab#1{#1}\fi

\bibitem[{Batygin(2015)}]{BATYGIN_2015}
Batygin, K. 2015, Monthly Notices of the Royal Astronomical Society, 451, 2589

\bibitem[{Batygin \& Morbidelli(2013)}]{BATYGIN-MORBIDELLI_2013}
Batygin, K. \& Morbidelli, A. 2013, \aj, 145, 1

\bibitem[{{Bou{\'e}} {et~al.}(2016){Bou{\'e}}, {Correia}, \&
  {Laskar}}]{BOUE-etal_2016}
{Bou{\'e}}, G., {Correia}, A. C.~M., \& {Laskar}, J. 2016, Celestial Mechanics
  and Dynamical Astronomy, 126, 31

\bibitem[{{Bou{\'e}} {et~al.}(2019){Bou{\'e}}, {Correia}, \&
  {Laskar}}]{BOUE-etal_2019}
{Bou{\'e}}, G., {Correia}, A.~C.~M., \& {Laskar}, J. 2019, in EAS Publications
  Series, Vol.~82, EAS Publications Series, 91--98

\bibitem[{{Bou{\'e}} \& {Efroimsky}(2019)}]{BOUE-EFROIMSKY_2019}
{Bou{\'e}}, G. \& {Efroimsky}, M. 2019, Celestial Mechanics and Dynamical
  Astronomy, 131, 30

\bibitem[{{Burkart} {et~al.}(2014){Burkart}, {Quataert}, \&
  {Arras}}]{BURKART-etal_2014}
{Burkart}, J., {Quataert}, E., \& {Arras}, P. 2014, \mnras, 443, 2957

\bibitem[{{Canup} \& {Ward}(2002)}]{CANUP-WARD_2002}
{Canup}, R.~M. \& {Ward}, W.~R. 2002, \aj, 124, 3404

\bibitem[{{Cassen} {et~al.}(1979){Cassen}, {Reynolds}, \&
  {Peale}}]{CASSEN-etal_1979}
{Cassen}, P., {Reynolds}, R.~T., \& {Peale}, S.~J. 1979, \grl, 6, 731

\bibitem[{Celletti {et~al.}(2019)Celletti, Paita, \&
  Pucacco}]{CELLETTI-etal_2019}
Celletti, A., Paita, F., \& Pucacco, G. 2019, Chaos, 29

\bibitem[{{Charalambous} {et~al.}(2018){Charalambous}, {Martí}, {Beaugé}, \&
  {Ramos}}]{CHARALAMBOUS-etal_2018}
{Charalambous}, C., {Martí}, J.~G., {Beaugé}, C., \& {Ramos}, X.~S. 2018,
  \mnras, 477, 1414

\bibitem[{{Darwin}(1880)}]{DARWIN_1880}
{Darwin}, G.~H. 1880, Philosophical Transactions of the Royal Society of
  London, Series I, 171, 713

\bibitem[{{de Sitter}(1909)}]{DESITTER_1909}
{de Sitter}, W. 1909, Proceedings of the Royal Netherlands Academy of Arts and
  Science, 11, 682

\bibitem[{{Deienno} {et~al.}(2014){Deienno}, {Nesvorný}, {Vokrouhlický}, \&
  {Yokoyama}}]{DEIENNO-etal_2014}
{Deienno}, R., {Nesvorný}, D., {Vokrouhlický}, D., \& {Yokoyama}, T. 2014,
  \aj, 148, 25

\bibitem[{{Dirkx} {et~al.}(2017){Dirkx}, {Gurvits}, {Lainey}, {Lari}, {Milani},
  {Cim{\`o}}, {Bocanegra-Bahamon}, \& {Visser}}]{DIRKX-etal_2017}
{Dirkx}, D., {Gurvits}, L.~I., {Lainey}, V., {et~al.} 2017, \planss, 147, 14

\bibitem[{{Efroimsky} \& {Lainey}(2007)}]{EFROIMSKY-LAINEY_2007}
{Efroimsky}, M. \& {Lainey}, V. 2007, Journal of Geophysical Research: Planets,
  112, E12003

\bibitem[{{Efroimsky} \& {Makarov}(2013)}]{EFROIMSKY-MAKAROV_2013}
{Efroimsky}, M. \& {Makarov}, V.~V. 2013, \apj, 764, 26

\bibitem[{{Everhart}(1985)}]{EVERHART_1985}
{Everhart}, E. 1985, in IAU Colloq. 83: Dynamics of Comets: Their Origin and
  Evolution, ed. A.~{Carusi} \& G.~B. {Valsecchi}, Vol. 115, 185

\bibitem[{{Ferraz-Mello}(2013)}]{FERRAZMELLO_2013}
{Ferraz-Mello}, S. 2013, Celestial Mechanics and Dynamical Astronomy, 116, 109

\bibitem[{{Ferraz-Mello} {et~al.}(2006){Ferraz-Mello}, {Michtchenko}, \&
  {Beaug{\'e}}}]{FERRAZMELLO-etal_2006}
{Ferraz-Mello}, S., {Michtchenko}, T.~A., \& {Beaug{\'e}}, C. 2006, in Chaotic
  Worlds: from Order to Disorder in Gravitational N-Body Dynamical Systems.,
  ed. B.~A. {Steves}, A.~J. {Maciejewski}, \& M.~{Hendry} (Springer
  Netherlands), 255--288

\bibitem[{{Ferraz-Mello} {et~al.}(2008){Ferraz-Mello}, {Rodriguez}, \&
  {Hussman}}]{FERRAZMELLO-etal_2008}
{Ferraz-Mello}, S., {Rodriguez}, A., \& {Hussman}, H. 2008, Celestial Mechanics
  and Dynamical Astronomy, 101, 171

\bibitem[{{Frouard} {et~al.}(2011){Frouard}, {Vienne}, \&
  {Fouchard}}]{FROUARD-etal_2011}
{Frouard}, J., {Vienne}, A., \& {Fouchard}, M. 2011, \aap, 532, A44

\bibitem[{Fuller {et~al.}(2016)Fuller, Luan, \& Quataert}]{FULLER-etal_2016}
Fuller, J., Luan, J., \& Quataert, E. 2016, \mnras, 458, 3867

\bibitem[{{Gallardo} {et~al.}(2016){Gallardo}, {Coito}, \&
  {Badano}}]{GALLARDO-etal_2016}
{Gallardo}, T., {Coito}, L., \& {Badano}, L. 2016, \icarus, 274, 83

\bibitem[{{Goldreich}(1965)}]{GOLDREICH_1965}
{Goldreich}, P. 1965, \mnras, 130, 159

\bibitem[{{Goldreich} \& {Soter}(1966)}]{GOLDREICH-SOTER_1966}
{Goldreich}, P. \& {Soter}, S. 1966, \icarus, 5, 375

\bibitem[{{Greenberg}(1973)}]{GREENBERG_1973}
{Greenberg}, R. 1973, \aj, 78, 338

\bibitem[{{Greenberg}(1982)}]{GREENBERG_1982}
{Greenberg}, R. 1982, in Satellites of Jupiter., ed. D.~{Morrison} (University
  of Arizona Press), 65--92

\bibitem[{{Greenberg}(1987)}]{GREENBERG_1987}
{Greenberg}, R. 1987, \icarus, 70, 334

\bibitem[{{Hussmann} \& {Spohn}(2004)}]{HUSSMANN-SPOHN_2004}
{Hussmann}, H. \& {Spohn}, T. 2004, \icarus, 171, 391

\bibitem[{{Iess} {et~al.}(2018){Iess}, {Folkner}, {Durante}, {Parisi}, {Kaspi},
  {Galanti}, {Guillot}, {Hubbard}, {Stevenson}, {Anderson}, {Buccino},
  {Casajus}, {Milani}, {Park}, {Racioppa}, {Serra}, {Tortora}, {Zannoni},
  {Cao}, {Helled}, {Lunine}, {Miguel}, {Militzer}, {Wahl}, {Connerney},
  {Levin}, \& {Bolton}}]{IESS-etal_2018}
{Iess}, L., {Folkner}, W.~M., {Durante}, D., {et~al.} 2018, \nat, 555, 220

\bibitem[{{Kaula}(1964)}]{KAULA_1964}
{Kaula}, W.~M. 1964, {Reviews of Geophysics}, 2, 661

\bibitem[{{Lainey} {et~al.}(2009){Lainey}, {Arlot}, {Karatekin}, \& {van
  Hoolst}}]{LAINEY-etal_2009}
{Lainey}, V., {Arlot}, J.-E., {Karatekin}, {\"O}., \& {van Hoolst}, T. 2009,
  \nat, 459, 957

\bibitem[{{Lainey} {et~al.}(2004{\natexlab{a}}){Lainey}, {Arlot}, \&
  {Vienne}}]{LAINEY-etal_2004b}
{Lainey}, V., {Arlot}, J.~E., \& {Vienne}, A. 2004{\natexlab{a}}, \aap, 427,
  371

\bibitem[{{Lainey} {et~al.}(2004{\natexlab{b}}){Lainey}, {Duriez}, \&
  {Vienne}}]{LAINEY-etal_2004}
{Lainey}, V., {Duriez}, L., \& {Vienne}, A. 2004{\natexlab{b}}, \aap, 420, 1171

\bibitem[{{Lainey} {et~al.}(2017){Lainey}, {Jacobson}, {Tajeddine}, {Cooper},
  {Murray}, {Robert}, {Tobie}, {Guillot}, {Mathis}, F., J., {Arlot}, {De
  Cuyper}, {Dehant}, {Pascu}, {Thuillot}, {Le Poncin-Lafitte}, \&
  {Zahn}}]{LAINEY-etal_2017}
{Lainey}, V., {Jacobson}, R.~A., {Tajeddine}, R., {et~al.} 2017, \icarus, 281,
  286

\bibitem[{{Lari}(2018)}]{LARI_2018}
{Lari}, G. 2018, Celestial Mechanics and Dynamical Astronomy, 130, 50

\bibitem[{{Lari} \& {Milani}(2019)}]{LARI-MILANI_2019}
{Lari}, G. \& {Milani}, A. 2019, \planss, 176, 104679

\bibitem[{{Laskar}(1990)}]{LASKAR_1990}
{Laskar}, J. 1990, \icarus, 88, 266

\bibitem[{{Laskar}(2005)}]{LASKAR_2005}
{Laskar}, J. 2005, in {Hamiltonian Systems and Fourier Analysis: New Prospects
  for Gravitational Dynamics}, ed. D.~Benest, C.~Froeschle, \& E.~Lega
  (Cambridge Scientific Publishers)

\bibitem[{{Laskar} \& {Bou{\'e}}(2010)}]{LASKAR-BOUE_2010}
{Laskar}, J. \& {Bou{\'e}}, G. 2010, \aap, 522, A60

\bibitem[{{Laskar} \& {Robutel}(1993)}]{LASKAR-ROBUTEL_1993}
{Laskar}, J. \& {Robutel}, P. 1993, \nat, 361, 608

\bibitem[{{Laskar} \& {Robutel}(1995)}]{LASKAR-ROBUTEL_1995}
{Laskar}, J. \& {Robutel}, P. 1995, Celestial Mechanics and Dynamical
  Astronomy, 62, 193

\bibitem[{{Le Maistre} {et~al.}(2016){Le Maistre}, {Folkner}, {Jacobson}, \&
  {Serra}}]{LEMAISTRE-etal_2016}
{Le Maistre}, S., {Folkner}, W.~M., {Jacobson}, R.~A., \& {Serra}, D. 2016,
  \planss, 126, 78

\bibitem[{{Love}(1909)}]{LOVE_1909}
{Love}, A.~E.~H. 1909, Proceedings of the Royal Society of London, Series A,
  82, 73

\bibitem[{{MacDonald}(1964)}]{MACDONALD_1964}
{MacDonald}, G. J.~F. 1964, Reviews of Geophysics and Space Physics, 2, 467

\bibitem[{{Malhotra}(1991)}]{MALHOTRA_1991}
{Malhotra}, R. 1991, \icarus, 94, 399

\bibitem[{{Meyer} \& {Wisdom}(2008)}]{MEYER-WISDOM_2008}
{Meyer}, J. \& {Wisdom}, J. 2008, \icarus, 193, 213

\bibitem[{{Mignard}(1979)}]{MIGNARD_1979}
{Mignard}, F. 1979, Moon and Planets, 20, 301

\bibitem[{{Murray} \& {Dermott}(2000)}]{MURRAY-DERMOTT_2000}
{Murray}, C.~D. \& {Dermott}, S.~F. 2000, {Solar System Dynamics} (Cambridge
  University Press)

\bibitem[{{Musotto} {et~al.}(2002){Musotto}, {Varadi}, {Moore}, \&
  {Schubert}}]{MUSOTTO-etal_2002}
{Musotto}, S., {Varadi}, F., {Moore}, W., \& {Schubert}, G. 2002, \icarus, 159,
  500

\bibitem[{{N{\'e}ron de Surgy} \& {Laskar}(1997)}]{NERONDESURGY-LASKAR_1997}
{N{\'e}ron de Surgy}, O. \& {Laskar}, J. 1997, \aap, 318, 975

\bibitem[{{Nesvorn{\'y}} \& {Morbidelli}(1998)}]{NESVORNY-MORBIDELLI_1998}
{Nesvorn{\'y}}, D. \& {Morbidelli}, A. 1998, Celestial Mechanics and Dynamical
  Astronomy, 71, 243

\bibitem[{{Noyelles} \& {Vienne}(2007)}]{NOYELLES-VIENNE_2007}
{Noyelles}, B. \& {Vienne}, A. 2007, \icarus, 190, 594–607

\bibitem[{{Ojakangas} \& {Stevenson}(1986)}]{OJAKANGAS-STEVENSON_1986}
{Ojakangas}, G.~W. \& {Stevenson}, D.~J. 1986, \icarus, 66, 341

\bibitem[{{Peale} \& {Cassen}(1978)}]{PEALE-CASSEN_1978}
{Peale}, S.~J. \& {Cassen}, P. 1978, \icarus, 36, 245

\bibitem[{{Peale} {et~al.}(1979){Peale}, {Cassen}, \&
  {Reynolds}}]{PEALE-etal_1979}
{Peale}, S.~J., {Cassen}, P., \& {Reynolds}, R.~T. 1979, Science, 203, 892

\bibitem[{{Peale} \& {Lee}(2002)}]{PEALE-LEE_2002}
{Peale}, S.~J. \& {Lee}, M.~H. 2002, Science, 298, 593

\bibitem[{Pichierri {et~al.}(2019)Pichierri, Batygin, \&
  Morbidelli}]{PICHIERRI-etal_2019}
Pichierri, G., Batygin, K., \& Morbidelli, A. 2019, A\&A, 625, A7

\bibitem[{{Poincar{\'e}}(1896)}]{POINCARE_1896}
{Poincar{\'e}}, H. 1896, in Comptes Rendus des s{\'e}ances de l'acad{\'e}mie
  des sciences, Vol. 123, 1031--1035

\bibitem[{{Polycarpe} {et~al.}(2018){Polycarpe}, {Saillenfest}, {Lainey},
  {Vienne}, {Noyelles}, \& {Rambaux}}]{POLYCARPE-etal_2018}
{Polycarpe}, W., {Saillenfest}, M., {Lainey}, V., {et~al.} 2018, \aap, 619,
  A133

\bibitem[{{Rein} \& {Spiegel}(2015)}]{REIN-SPIEGEL_2015}
{Rein}, H. \& {Spiegel}, D.~S. 2015, \mnras, 446, 1424

\bibitem[{{Saillenfest} {et~al.}(2019){Saillenfest}, {Laskar}, \&
  {Bou{\'e}}}]{SAILLENFEST-etal_2019}
{Saillenfest}, M., {Laskar}, J., \& {Bou{\'e}}, G. 2019, \aap, 623, A4

\bibitem[{{Serra} {et~al.}(2019){Serra}, {Lari}, {Tommei}, {Durante},
  {Gomez Casajus}, {Notaro}, {Zannoni}, {Iess}, {Tortora}, \&
  {Bolton}}]{SERRA-etal_2019}
{Serra}, D., {Lari}, G., {Tommei}, G., {et~al.} 2019, \mnras, 490, 766

\bibitem[{Showman \& Malhotra(1997)}]{SHOWMAN-MALHOTRA_1997}
Showman, A.~P. \& Malhotra, R. 1997, \icarus, 127, 93

\bibitem[{{Sinclair}(1972)}]{SINCLAIR_1972}
{Sinclair}, A.~T. 1972, \mnras, 160, 169

\bibitem[{{Sinclair}(1975)}]{SINCLAIR_1975}
{Sinclair}, A.~T. 1975, Celestial Mechanics, 12, 89

\bibitem[{{Singer}(1968)}]{SINGER_1968}
{Singer}, S.~F. 1968, Geophysical Journal International, 15, 205

\bibitem[{{Souillart}(1880)}]{SOUILLARD_1880}
{Souillart}, M. 1880, \memras, 45, 1

\bibitem[{{Tittemore}(1990)}]{TITTEMORE_1990}
{Tittemore}, W.~C. 1990, Science, 250, 263

\bibitem[{{Tittemore} \& {Wisdom}(1988)}]{TITTEMORE-WISDOM_1988}
{Tittemore}, W.~C. \& {Wisdom}, J. 1988, \icarus, 74, 172

\bibitem[{{Tittemore} \& {Wisdom}(1990)}]{TITTEMORE-WISDOM_1990}
{Tittemore}, W.~C. \& {Wisdom}, J. 1990, \icarus, 85, 394

\bibitem[{{Ward} \& {Canup}(2006)}]{WARD-CANUP_2006}
{Ward}, W.~R. \& {Canup}, R.~M. 2006, \apj, 640, L91

\bibitem[{Yoder(1979)}]{YODER_1979}
Yoder, C.~F. 1979, \nat, 279, 767

\bibitem[{{Yoder} \& {Peale}(1981)}]{YODER-PEALE_1981}
{Yoder}, C.~F. \& {Peale}, S.~J. 1981, \icarus, 47, 1

\end{thebibliography}


   \appendix
   

\section{Building the Hamiltonian function}\label{asec:ham}
In this section, we summarize the method used to obtain the averaged Hamiltonian model described in Sect.~\ref{sec:model}. The basic procedure is the same as in \cite{LARI_2018}, but the noninertial nature of the reference frame requires a specific treatment.

We consider a set of bodies $i=0,1..,N$ with masses $m_i$ and positions $\mathbf{x}_i$ measured in an inertial reference system. In our case, the index $0$ is Jupiter, and the indexes $1$ to $N=4$ are the Galilean satellites. Their equations of motion are
\begin{equation}
   m_i\ddot{\mathbf{x}}_i = \mathbf{F}_i\ \ \forall\ i=0,1...N\,,
\end{equation}
where $\mathbf{F}_i$ is the force applied to body $i$. We introduce the barycentric coordinates $\mathbf{y}_i$ such that
\begin{equation}
   \sum_{i=0}^Nm_i\mathbf{y}_i = \mathbf{0}
   \hspace{0.5cm}\text{and}\hspace{0.5cm}
   \mathbf{x}_i = \mathbf{x}_\text{G} + \mathbf{y}_i \ \ \forall\ i=0,1...N\,,
\end{equation}
by definition. The barycenter of the system is located in $\mathbf{x}_\mathrm{G}$ in the inertial reference system. It undergoes a nonzero acceleration, mainly due to the gravitational attraction of the Sun. Therefore, the equations of motion become
\begin{equation}\label{eq:motion}
    m_i\ddot{\mathbf{y}}_i = \mathbf{F}_i - m_i\ddot{\mathbf{x}}_\text{G}\ \ \forall\ i=0,1...N\,.
\end{equation}
From the definition of the barycenter, the dynamics of one body (and in particular, Jupiter) can also be expressed as
\begin{equation}\label{eq:accJup}
   m_0\ddot{\mathbf{y}}_0 = -\sum_{i=1}^N\mathbf{F}_i + \ddot{\mathbf{x}}_\text{G}\sum_{i=1}^Nm_i \,.
\end{equation}
Taking into account the mutual attraction between the bodies, the nonsphericity of Jupiter, and the attraction of the Sun, the force applied to a satellite $i=1,2...N$ is
\begin{equation}
   \mathbf{F}_i = -\sum_{\substack{k=0\\k\neq i}}^N\frac{\mathcal{G}m_im_k}{\|\mathbf{y}_i-\mathbf{y}_k\|^3}(\mathbf{y}_i-\mathbf{y}_k) + \mathbf{F}_i^{\text{J}} - \frac{\mathcal{G}m_im_\odot}{\|\mathbf{y}_i-\mathbf{y}_\odot\|^3}(\mathbf{y}_i-\mathbf{y}_\odot)\,,
\end{equation}
where $m_\odot$ is the mass of the Sun and $\mathbf{y}_\odot$ its position with respect to the barycenter of bodies $0,1...N$. The vector $\mathbf{F}_i^\text{J}$ is the force applied to the $i$th satellite because of the non-sphericity of Jupiter; it only depends on $\mathbf{y}_i-\mathbf{y}_0$. By summation, we obtain Jupiter's equation of motion through Eq.~\eqref{eq:accJup}. Assuming that the vector $\mathbf{x}_\mathrm{G}$ is a known function of time $t$, the equations of motion can be established from the Lagrangian function
\begin{equation}\label{eq:Lag}
   \mathcal{L} = \sum_{i=0}^N\frac{1}{2}m_i\|\dot{\mathbf{y}}_i\|^2 - U(\mathbf{y}_0,\mathbf{y}_1...\mathbf{y}_N,t)\,,
\end{equation}
where
\begin{equation}
   \begin{aligned}
      U &= -\sum_{0\leqslant i<k\leqslant N}\frac{\mathcal{G}m_im_k}{\|\mathbf{y}_i-\mathbf{y}_k\|}
      + \sum_{i=1}^NU_i^{\text{J}} - \sum_{i=1}^N\frac{\mathcal{G}m_im_\odot}{\|\mathbf{y}_i-\mathbf{y}_\odot\|} \\
      &+ \ddot{\mathbf{x}}_\text{G}\cdot\sum_{i=1}^Nm_i(\mathbf{y}_i-\mathbf{y}_0) \,,
   \end{aligned}
\end{equation}
and
\begin{equation}
   \mathbf{F}_i^\text{J} = -\frac{\partial U_i^\text{J}}{\partial\mathbf{y}_i} \ \ \forall\ i=1,2...N\,.
\end{equation}
The potential energy $U_i^\text{J}$ is only function of $\mathbf{y}_i-\mathbf{y}_0$. By applying the Lagrange equations to Eq.~\eqref{eq:Lag}, we exactly retrieve Eq.~\eqref{eq:motion} for bodies $1$ to $N$. For body $0$, we retrieve Eq.~\eqref{eq:accJup} by neglecting terms of order $\|\mathbf{y}_0\|/\|\mathbf{y}_\odot\|$, which is about $10^{-7}$ for Jupiter and its satellites.

We now consider the positions $\mathbf{z}_i$ of the bodies in a frame with the third axis oriented along the spin of Jupiter and the first axis directed towards its instantaneous equinox. This reference frame rotates with respect to the previous one with a rotation vector $\mathbf{\Theta}(t)$ due to motion of the planet's spin-axis and the variations of its orbit. The Varignon-Bour formula gives the following composition laws:
\begin{equation}
   \left\{
   \begin{aligned}
      \mathbf{y}_i &= \mathbf{z}_i \\
      \dot{\mathbf{y}}_i &= \dot{\mathbf{z}}_i + \mathbf{\Theta}\times\mathbf{z}_i
   \end{aligned}
   \right.\ \ \forall\ i=0,1...N\,,
\end{equation}
where $\dot{\mathbf{z}}_i$ is the time derivative of $\mathbf{z}_i$ as measured in the rotating frame. In the new coordinates, the Lagrangian in Eq.~\eqref{eq:Lag} becomes
\begin{equation}
   \mathcal{L} = \sum_{i=0}^N\frac{1}{2}m_i\|\dot{\mathbf{z}}_i + \mathbf{\Theta}\times\mathbf{z}_i\|^2 - U(\mathbf{z}_0,\mathbf{z}_1...\mathbf{z}_N,t)\,.
\end{equation}
We now introduce the momentum $\mathbf{Z}_i$ conjugate to $\mathbf{z}_i$, defined by
\begin{equation}
   \mathbf{Z}_i = \frac{\partial\mathcal{L}}{\partial\dot{\mathbf{z}}_i} = m_i(\dot{\mathbf{z}}_i + \mathbf{\Theta}\times\mathbf{z}_i) = m_i\dot{\mathbf{y}}_i\ \ \forall\ i=0,1...N\,.
\end{equation}
This leads to the following Hamiltonian function:
\begin{equation}
   \begin{aligned}
      \mathcal{H} &= \sum_{i=0}^N\mathbf{Z}_i\cdot\dot{\mathbf{z}}_i - \mathcal{L} \\
      &= \sum_{i=0}^N\frac{1}{2}\frac{\|\mathbf{Z}_i\|^2}{m_i} + U(\mathbf{z}_0,\mathbf{z}_1...\mathbf{z}_N,t) - \mathbf{\Theta}\cdot\sum_{i=0}^N\mathbf{z}_i\times\mathbf{Z}_i \,.
   \end{aligned}
\end{equation}
By writing down Hamilton's equations for $\mathbf{Z}_i$ and $\mathbf{z}_i$, we retrieve the classical formula of the inertial forces produced in an accelerated rotating frame.

Finally, we switch to Jovicentric canonical coordinates following the original idea of \cite{POINCARE_1896} applied for instance by \cite{LASKAR-ROBUTEL_1995} or \cite{FERRAZMELLO-etal_2006}. An elegant variant has been found by Gwena{\"e}l Bou{\'e} (private communication), leading to the coordinates
\begin{equation}
  \left\{
  \begin{aligned}
     \mathbf{r}_0 &= \sum_{k=0}^N\frac{m_k}{M_\text{tot}}\mathbf{z}_k \,,\\
     \mathbf{r}_i &= \mathbf{z}_i - \mathbf{z}_0\ \ \forall i = 1,2...N\,,
  \end{aligned}
  \right.
\end{equation}
and conjugate momenta
\begin{equation}
  \left\{
  \begin{aligned}
     \mathbf{p}_0 &= \sum_{k=0}^N\mathbf{Z}_k\,, \\
     \mathbf{p}_i &= \mathbf{Z}_i - \frac{m_i}{M_\text{tot}}\sum_{k=0}^N\mathbf{Z}_k\ \ \forall i = 1,2...N\,,
  \end{aligned}
  \right.
\end{equation}
where
\begin{equation}
M_\mathrm{tot} \equiv \sum_{j=0}^Nm_j \,.
\end{equation}
The coordinates $\mathbf{r}_1$ to $\mathbf{r}_N$ are the Jovicentric position vectors of the satellites, and $\mathbf{r}_0$ is the location of the barycenter of the planet and its satellites. In order to express the Hamiltonian function in the new coordinates, we note that
\begin{equation}
   \sum_{i=0}^N\frac{1}{2}\frac{\|\mathbf{Z}_i\|^2}{m_i} = \frac{1}{2}\frac{\|\mathbf{p}_0\|^2}{M_\text{tot}} + \sum_{i=1}^N\frac{1}{2}\frac{\|\mathbf{p}_i\|^2}{\beta_i} + \sum_{1\leqslant i<j\leqslant N}\frac{\mathbf{p}_i\cdot\mathbf{p}_k}{m_0}\,,
\end{equation}
where $\beta_i=m_0m_i/(m_0+m_i)$, and that
\begin{equation}
   \sum_{i=0}^N\mathbf{z}_i\times\mathbf{Z}_i = \sum_{i=0}^N\mathbf{r}_i\times\mathbf{p}_i \,.
\end{equation}
Therefore, after having introduced the Jovicentric position of the Sun $\mathbf{r}_\odot = \mathbf{y}_\odot - \mathbf{y}_0$ supposed to be a known function of time, the coordinates $\mathbf{r}_0$ and $\mathbf{p}_0$ appear as completely isolated in the Hamiltonian function (whatever their value). Accordingly the corresponding terms can be dropped. The final form of the Hamiltonian function is then $\mathcal{H} = \mathcal{H}_0 + \varepsilon\mathcal{H}_1$, in which $\varepsilon\mathcal{H}_1 = \mathcal{H}_\text{J} + \mathcal{H}_\text{M} + \mathcal{H}_\odot + \mathcal{H}_\text{I}$, with
\begin{equation}\label{eq:Htot}
   \begin{aligned}
      \mathcal{H}_0 &= \sum_{i=1}^N\left(\frac{\|\mathbf{p}_i\|^2}{2\beta_i} - \frac{\mu_i\beta_i}{\|\mathbf{r}_i\|}\right) \,,\\
      \mathcal{H}_\mathrm{J} &= \sum_{i=1}^NU_i^{\text{J}}(\mathbf{r}_i) \,,\\
      \mathcal{H}_\mathrm{M} &= - \sum_{1\leqslant i<k\leqslant N}\left(\frac{\mathcal{G}m_im_k}{\|\mathbf{r}_i-\mathbf{r}_k\|} - \frac{\mathbf{p}_i\cdot\mathbf{p}_k}{m_0}\right) \,,\\
      \mathcal{H}_\odot &= - \sum_{i=1}^N\frac{\mathcal{G}m_im_\odot}{\|\mathbf{r}_i-\mathbf{r}_\odot\|} + \ddot{\mathbf{x}}_\text{G}\cdot\sum_{i=1}^Nm_i\mathbf{r}_i \,,\\
      \mathcal{H}_\mathrm{I} &= - \mathbf{\Theta}\cdot\sum_{i=1}^N\mathbf{r}_i\times\mathbf{p}_i \,,
   \end{aligned}
\end{equation}
where $\mu_i = \mathcal{G}(m_0+m_i)$. The dominant part $\mathcal{H}_0$ is a sum of unperturbed Kepler problems with mass $\beta_i$ and $\mu$-parameter $\mu_i$. In order to follow a perturbative approach, we then replace $\mathbf{r}_i$ and $\mathbf{p}_i$ by coordinates that are ``action-angle'' for $\mathcal{H}_0$, such as the Delaunay canonical coordinates given in Eq.~\eqref{eq:Delaunay}. In the context of our secular theory, each term is eventually averaged over the short-period terms and expanded into suitable series. The explicit expression of each part is described in Sect.~\ref{sec:model}.

The solar term $\mathcal{H}_\odot$ deserves further clarifications. In Eq.~\eqref{eq:Htot}, we chose to include the terms involving $\ddot{\mathbf{x}}_\mathrm{G}$ into the definition of $\mathcal{H}_\odot$ instead of putting them into the inertial part $\mathcal{H}_\mathrm{I}$. Indeed, the acceleration of the barycenter of Jupiter and its satellites is largely dominated by the attraction of the Sun; the instantaneous attraction from the other planets of the Solar System is neglected. This leads to the classic ``indirect'' potential in the Hamiltonian\footnote{ $\ddot{\mathbf{x}}_\mathrm{G}$ as a function of time could be taken from the ephemerides, as we do for $\mathbf{\Theta}$ (see Appendix~\ref{asec:QPS}). However, this would introduce an unnecessary computational complexity.}:
\begin{equation}
   \ddot{\mathbf{x}}_\mathrm{G} \approx \frac{\mathcal{G}m_\odot}{\|\mathbf{y}_\odot\|^3}\mathbf{y}_\odot =  \frac{\mathcal{G}m_\odot}{\|\mathbf{r}_\odot\|^3}\mathbf{r}_\odot + \mathcal{O}\left(\frac{\|\mathbf{y}_0\|}{\|\mathbf{r}_\odot\|}\right)\,.
\end{equation}
When expanding $\mathcal{H}_\odot$ in Legendre polynomials, this term cancels exactly the first order in $a_i/a_\odot$. This is why Eq.~\eqref{eq:Hsun} starts at second order. Then, the Sun's orbital elements can be gathered into the coefficients $C_1^\odot$ to $C_9^\odot$ of Eq.~\eqref{eq:Hsun}. These coefficients are
\begin{equation}
   \begin{aligned}
      C_1^\odot &= \frac{3}{16}\sin^2I_\odot\Bigg(-17e_\odot^2\cos(4\lambda_\odot-2\varpi_\odot) - 7e_\odot\cos(3\lambda_\odot-\varpi_\odot) \\
      &\hspace{2cm}
      + e_\odot\cos( \lambda_\odot+\varpi_\odot) + (5e_\odot^2-2)\cos(2\lambda_\odot)\Bigg) \\
      &- \frac{1}{16}(3\cos^2I_\odot - 1)\Bigg(9e_\odot^2\cos(2\lambda_\odot - 2\varpi_\odot) \\
      &\hspace{2.8cm}
      + 6e_\odot\cos(\lambda_\odot-\varpi_\odot) + 3e_\odot^2 + 2\Bigg) \\
      C_2^\odot &= - \frac{3}{16}\Bigg(3\cos^2I_\odot + 3\sin^2I_\odot\cos(2\lambda_\odot) - 1\Bigg) \\
      C_3^\odot &= - \frac{3}{4}\Bigg(\sin^2I_\odot +(\cos^2I_\odot + 1)\cos(2\lambda_\odot)\Bigg) \\
      C_4^\odot &= - \frac{3}{2}\cos I_\odot\sin(2\lambda_\odot) \\
      C_5^\odot &= \frac{3}{4}\cos I_\odot\sin I_\odot\Bigg(7e_\odot \cos(3\lambda_\odot-\varpi_\odot) - 6e_\odot \cos(\lambda_\odot-\varpi_\odot) \\
      &\hspace{2.4cm}
      - e_\odot\cos(\lambda_\odot+\varpi_\odot) + 2\cos(2\lambda_\odot) - 2\Bigg) \\
      C_6^\odot &= \frac{3}{4}\sin I_\odot\Bigg(7e_\odot\sin(3\lambda_\odot-\varpi_\odot) - e_\odot\sin(\lambda_\odot+\varpi_\odot) + 2\sin(2\lambda_\odot)\Bigg) \\
      C_7^\odot &= \frac{15}{64}\Bigg(5\sin^2I_\odot\cos(3\lambda_\odot) + (5\cos^2I_\odot - 1)\cos\lambda_\odot\Bigg) \\
      C_8^\odot &= \frac{15}{64}\cos I_\odot\Bigg(5\sin^2I_\odot\sin(3\lambda_\odot) + (15\cos^2I_\odot - 11)\sin\lambda_\odot\Bigg) \\
      C_9^\odot &= -\frac{3}{512}\Bigg(20(7\cos^2I_\odot - 1)\sin^2I_\odot\cos(2\lambda_\odot) \\
      &\hspace{0.9cm}
      + 35\sin^4I_\odot\cos(4\lambda_\odot) + 3(35\cos^4I_\odot - 30\cos^2I_\odot + 3)\Bigg) \\
   \end{aligned}
\end{equation}
in our reference frame (where $\Omega_\odot=0$ by definition). In these expressions, $e_\odot$ is the eccentricity of the Sun, $I_\odot$ its inclination, $\varpi_\odot$ its longitude of perihelion and $\lambda_\odot$ its mean longitude. Each of these elements, as well as the semi-major axis $a_\odot$ also appearing in Eq.~\eqref{eq:Hsun}, vary with time as described in Appendix~\ref{asec:QPS}.
   
   
\section{Orbital and rotational evolution of Jupiter}\label{asec:QPS}
The orbital perturbations taken into account in our model of the Galilean satellites are summarized in Eq.~\eqref{eq:eH1}. In order to compute the Sun's varying orbital elements appearing in $\mathcal{H}_\odot$ and the inertial terms $\mathcal{H}_\mathrm{I}$, we need to have a previous knowledge of the orbital and rotational evolution of Jupiter in the Solar System. We give below the solutions that we use and describe how they were obtained.

\subsection{Orbital solution}
We need an orbital solution for Jupiter that would be valid on a gigayear timescale. This is well beyond the timespan covered by ephemerides. Luckily, the orbital dynamics of the giant planets of the Solar System are (almost) integrable, and excellent solutions have been developed. We use the secular solution of \cite{LASKAR_1990}, obtained by multiplying the normalized proper modes $z_i^\bullet$ and $\zeta_i^\bullet$ (Tables VI and VII of \citealp{LASKAR_1990}) by  the matrix $\tilde{S}$ corresponding to the linear part of the solution (Table V of \citealp{LASKAR_1990}). In the series obtained, the terms with the same combination of frequencies are then merged together, resulting in 56 terms in eccentricity and 60 terms in inclination. However, this only forms the secular part of the orbital solution; the short-term component (i.e., the planets' orbital timescale) is slow compared to the motion of the Galilean satellites, so it must be included as well. In order to build a complete orbital solution, we subtracted the secular part from the 2000 yr timespan of the \texttt{INPOP17a} ephemerides\footnote{\texttt{https://www.imcce.fr/inpop}}, and we ran a frequency analysis (see e.g., \citealp{LASKAR_2005}) on the result. This gave the short-term part of the solution. Finally, the complete orbital solution was made by adding together the short-term and secular series obtained.

The orbital solution is expressed in the following variables:
\begin{equation}
   \begin{aligned}
      p &= \frac{n}{N} - 1 &=&& \sum_k P_k\cos(\omega_kt+\alpha_k^{(0)})\,,\\
      q &= i(\lambda - Nt- \lambda_0) &=&& i\sum_k Q_k\sin(\gamma_kt+\beta_k^{(0)})\,,\\
      z &= e\exp(i\varpi) &=&& \sum_k E_k\exp\big[i(\mu_kt+\theta_k^{(0)})\big]\,,\\
      \zeta &= \sin\frac{I}{2}\exp(i\Omega) &=&& \sum_k S_k\exp\big[i(\nu_kt+\phi_k^{(0)})\big] \,.
   \end{aligned}
\end{equation}
The quantities $z$ and $\zeta$ are complex numbers, whereas $p$ is real and $q$ is pure imaginary. In these expressions, $n$ is the mean motion of Jupiter, $\lambda$ its mean longitude, $e$ its eccentricity, $\varpi$ its longitude of perihelion, $I$ its inclination, and $\Omega$ its longitude of ascending node. The time is noted $t$. By virtue of trigonometric identities, moving Jupiter one step forward in time using the quasi-periodic decomposition only amounts to computing a few sums and products.

In Tables~\ref{tab:p} to \ref{tab:zeta}, we give the terms of the solution in the J2000 ecliptic and equinox reference frame, for amplitudes up to order $10^{-5}$. These terms contain contributions from all the planets of the Solar System, including in particular the great 2:5 Jupiter--Saturn inequality, which is known to play a role in the dynamics of several Jovian satellites \citep{FROUARD-etal_2011}.

\begin{table}
   \caption{Quasi-periodic decomposition of Jupiter's mean motion (variable $p$).}
   \label{tab:p}
   \vspace{-0.7cm}
   \begin{equation*}
      \begin{array}{rrrr}
      \hline
      \hline
      k & \omega_k\ (''\cdot\text{yr}^{-1}) & P_k\times 10^5 & \alpha_k^{(0)}\ (^\text{o}) \\
      \hline
  1 &   130520.10160 &    20 & 148.12 \\
  2 &   -21277.78083 &     9 &  71.07 \\
  3 &   195780.06735 &     9 & 133.11 \\
  4 &    -1387.39180 &     7 & 186.64 \\
  5 &   -86550.40389 &     6 & 139.88 \\
  6 &   -65261.39096 &     6 & 197.29 \\
  7 &  -261040.14054 &     4 & 241.93 \\
  8 &   151810.07834 &     3 & 203.76 \\
  9 &  1186720.95784 &     2 & 246.12 \\
 10 & -1997384.90488 &     2 &  32.35 \\
 11 &   326300.22618 &     2 & 103.34 \\
 12 &   217070.09240 &     2 & 187.88 \\
 13 &   -42579.92557 &     2 &  66.82 \\
 14 &   -22678.66367 &     2 & 182.96 \\
 15 &  -282334.09223 &     1 & 188.36 \\
      \hline
      \end{array}
   \end{equation*}
   \vspace{-0.5cm}
   \tablefoot{The phases $\alpha_k^{(0)}$ are given at time J2000.}
\end{table}

\begin{table}
   \caption{Quasi-periodic decomposition of Jupiter's mean longitude (variable $q$).}
   \label{tab:q}
   \vspace{-0.7cm}
   \begin{equation*}
      \begin{array}{rrrr}
      \hline
      \hline
      k & \gamma_k\ (''\cdot\text{yr}^{-1}) & Q_k\times 10^5 & \beta_k^{(0)}\ (^\text{o}) \\
      \hline
  1 &     1382.39672 &   565 & 173.33 \\
  2 &    21279.46165 &    62 & 285.69 \\
  3 &  -130520.09747 &    32 &  31.94 \\
  4 &   -65260.75362 &    24 &  16.88 \\
  5 &      740.73142 &    20 & 111.50 \\
  6 &   -86550.20151 &    13 & 316.81 \\
  7 &   195780.09376 &    12 & 132.75 \\
  8 &    -2146.66254 &     9 & 340.54 \\
  9 &    42565.96834 &     7 & 296.80 \\
 10 &   151810.10095 &     5 & 206.50 \\
 11 &   -22663.04452 &     5 &  14.91 \\
 12 &    43974.51084 &     5 &  98.00 \\
 13 &  -261040.17870 &     5 &  62.52 \\
 14 &     3182.71336 &     3 & 148.86 \\
 15 &   217070.21223 &     2 & 190.22 \\
 16 &  -326300.28181 &     2 &  77.50 \\
 17 &  -109248.95417 &     2 & 121.83 \\
 18 &  -107838.32524 &     2 & 259.67 \\
 19 &  1186720.95929 &     1 &  66.12 \\
 20 &  1997384.90244 &     1 & 147.64 \\
 21 &  -282334.67409 &     1 &   6.37 \\
 22 &    20350.19793 &     1 & 351.20 \\
      \hline
      \end{array}
   \end{equation*}
   \vspace{-0.5cm}
   \tablefoot{The phases $\beta_k^{(0)}$ are given at time J2000. The mean longitude is given by $\lambda=Nt+\lambda_0-iq$, where $N = 0.52969$~rad.yr$^{-1}$ and $\lambda_0 = 0.59946$~rad with the time $t$ measured from J2000.}
\end{table}

\begin{table}
   \caption{Quasi-periodic decomposition of Jupiter's eccentricity and longitude of perihelion (variable $z$).}
   \label{tab:z}
   \vspace{-0.7cm}
   \begin{equation*}
      \begin{array}{rrrr}
      \hline
      \hline
      k & \mu_k\ (''\cdot\text{yr}^{-1}) & E_k\times 10^5 & \theta_k^{(0)}\ (^\text{o}) \\
      \hline
  1 &        4.24882 &  4412 &  30.67 \\
  2 &       28.22069 &  1575 & 308.11 \\
  3 &        3.08952 &   180 & 121.36 \\
  4 &   -21263.65777 &    65 &  66.27 \\
  5 &       52.19257 &    52 &  45.55 \\
  6 &     1410.36662 &    38 & 116.79 \\
  7 &       27.06140 &    18 & 218.71 \\
  8 &       29.37998 &    18 & 217.53 \\
  9 &    22706.58543 &    13 & 172.92 \\
 10 &       28.86795 &    11 &  32.64 \\
 11 &   -86523.67052 &    11 &  81.91 \\
 12 &       27.57346 &     9 &  43.74 \\
 13 &    43995.78824 &     8 & 231.01 \\
 14 &   -42553.63044 &     6 &   9.94 \\
 15 &        5.40817 &     6 & 120.31 \\
 16 &        0.66708 &     6 &  73.98 \\
 17 &       53.35188 &     4 & 314.90 \\
 18 &  -151783.76249 &     4 &  97.31 \\
 19 &   109255.80241 &     3 & 214.22 \\
 20 &       76.16447 &     2 & 143.03 \\
 21 &       56.32774 &     2 &  95.77 \\
 22 &  -107813.70709 &     2 &  26.23 \\
 23 &  -217043.86396 &     2 & 112.41 \\
 24 &    87975.02083 &     1 & 115.18 \\
 25 &   239776.59923 &     1 &   2.29 \\
 26 &       51.03334 &     1 & 316.30 \\
 27 &        7.45592 &     1 &  20.24 \\
 28 &      -19.72306 &     1 & 293.24 \\
 29 &    21305.79949 &     1 & 356.32 \\
 30 &  1295977.39395 &     1 & 100.47 \\
 31 &   -22669.06392 &     1 & 253.27 \\
      \hline
      \end{array}
   \end{equation*}
   \vspace{-0.5cm}
   \tablefoot{The phases $\theta_k^{(0)}$ are given at time J2000.}
\end{table}

\begin{table}
   \caption{Quasi-periodic decomposition of Jupiter's inclination and longitude of ascending node (variable $\zeta$).}
   \label{tab:zeta}
   \vspace{-0.7cm}
   \begin{equation*}
      \begin{array}{rrrr}
      \hline
      \hline
      k & \nu_k\ (''\cdot\text{yr}^{-1}) & S_k\times 10^5 & \phi_k^{(0)}\ (^\text{o}) \\
      \hline
  1 &        0.00000 &  1377 & 107.59 \\
  2 &      -26.33023 &   315 & 307.29 \\
  3 &       -0.69189 &    58 &  23.96 \\
  4 &       -3.00557 &    48 & 140.33 \\
  5 &      -26.97744 &     2 & 222.98 \\
  6 &       -2.35835 &     2 &  44.74 \\
  7 &       82.77163 &     1 & 308.95 \\
  8 &       -1.84625 &     1 &  36.64 \\
  9 &       -5.61755 &     1 & 168.70 \\
      \end{array}
   \end{equation*}
   \vspace{-0.5cm}
   \tablefoot{The phases $\phi_k^{(0)}$ are given at time J2000.}
\end{table}

\subsection{Rotational solution}
The precession constant of Jupiter, which depends on its moments of inertia, is not perfectly known. As reported by \cite{WARD-CANUP_2006}, the spin axis of Jupiter is very close to the Cassini state 2 with the precession of Uranus' node (term $k=4$ of Table~\ref{tab:zeta}). For this reason, a small change of Jupiter's precession constant leads to quite different evolutions for the spin-axis, because it moves Jupiter closer or farther from this Cassini state.

Moreover, the precession constant of Jupiter also depends on the distance of its most massive satellites. Therefore, the tidal migration of the Galilean satellites could also lead the spin axis of Jupiter closer or farther from this Cassini state. This led \cite{WARD-CANUP_2006} to conjecture that Jupiter's spin axis has been attracted long term ago into this Cassini state due to dissipations, and that the current value of its precession constant is not $2.74 ''\cdot$yr$^{-1}$, as nominally predicted by the available data, but rather $2.94 ''\cdot$yr$^{-1}$ (which remains compatible with the uncertainties). This would put Jupiter just near the Cassini state 2 with the precession of Uranus' node.

The question of the value of Jupiter's precession constant and its update using modern spatial missions like Juno is very interesting (see e.g., \citealp{LEMAISTRE-etal_2016}), but it goes well beyond the scope of this paper. Here, we restrict our goal to avoiding to make the satellites' dynamics over-stable because of considering a fixed obliquity for Jupiter. Therefore, we need a realistic evolution for Jupiter's spin axis, but we do not pretend to model it in all its subtlety. We obtained such a solution by fixing the precession constant of Jupiter to its nominal value ($2.74 ''\cdot$yr$^{-1}$), and by performing a one-gigayear numerical integration of the secular rotational equations (see e.g., \citealp{LASKAR-ROBUTEL_1993,NERONDESURGY-LASKAR_1997}). To this end, we used the forcing from the secular part of the orbital solution given in Appendix~\ref{asec:QPS} (this method has been proved to give very good results for the planets of the Solar System, see \citealp{SAILLENFEST-etal_2019}). Then, the spin-axis solution was put under the form of a synthetic series, using a frequency analysis to the variable
\begin{equation}
   y = \sin\frac{\varepsilon}{2}\exp(i\psi) = \sum_k Y_k\cos(\eta_kt+\delta_k^{(0)})\,,
\end{equation}
where $\varepsilon$ is the obliquity of Jupiter and $\psi$ its precession angle. The spin-axis solution obtained is given in Table~\ref{tab:y} with amplitudes up to $10^{-5}$.

\begin{table}
   \caption{Quasi-periodic decomposition of Jupiter's obliquity and precession angle (variable $y$).}
   \label{tab:y}
   \vspace{-0.7cm}
   \begin{equation*}
      \begin{array}{rrrr}
      \hline
      \hline
      k & \eta_k\ (''\cdot\text{yr}^{-1}) & Y_k\times 10^5 & \delta_k^{(0)}\ (^\text{o}) \\
      \hline
  1 &    2.74657 &  2505 & 225.47 \\
  2 &    3.00557 &   551 & 219.67 \\
  3 &   26.33023 &   352 &  52.71 \\
  4 &    0.69189 &    20 & 156.04 \\
  5 &    2.48757 &     8 & 231.28 \\
  6 &    2.35835 &     7 & 135.24 \\
  7 &    3.11725 &     3 &  33.03 \\
  8 &    4.16482 &     3 & 308.44 \\
  9 &   26.97744 &     3 & 137.02 \\
 10 &    1.84625 &     3 & 142.36 \\
 11 &    5.61755 &     2 & 191.30 \\
 12 &  -82.77163 &     1 &  51.05 \\
      \hline
      \end{array}
   \end{equation*}
   \vspace{-0.5cm}
   \tablefoot{The phases $\delta_k^{(0)}$ are given at time J2000.}
\end{table}

\subsection{Inertial terms}
Once an orbital and rotational solution for Jupiter is known, the computation of the inertial term $\mathcal{H}_\mathrm{I}$ at any time is straightforward. As explained in Appendix~\ref{asec:ham}, the vector $\mathbf{\Theta}$ is the rotation velocity of our rotating reference frame (with the $z$ axis perpendicular to Jupiter's equator and the $x$ axis directed towards its equinox) measured in a nonrotating reference frame. For instance, the rotation matrix $R$ that converts the coordinates of a vector expressed in our reference frame towards the J2000 ecliptic and equinox reference frame is
\begin{equation}
   R = R_z(\Omega)R_x(I)R_z(-\Omega)R_z(-\psi)R_x(-\varepsilon)
,\end{equation}
where
\begin{equation}
   R_x(\alpha) = 
   \begin{pmatrix}
      1 & 0 & 0 \\
      0 & \cos\alpha & -\sin\alpha \\
      0 & \sin\alpha & \cos\alpha \\
   \end{pmatrix}
   \,,\hspace{0.2cm}
   R_z(\alpha) = 
   \begin{pmatrix}
      \cos\alpha & -\sin\alpha & 0\\
      \sin\alpha &  \cos\alpha & 0\\
      0 & 0 & 1
   \end{pmatrix}
   \,.
\end{equation}
The transformation $R$ can be considered as a single rotation of angle $\theta$ about an inclined axis. Writing $\mathbf{n}=(n_x,n_y,n_z)^\mathrm{T}$ the unitary vector that defines this axis, we have
\begin{equation}
   \mathbf{\Theta} = \dot{\theta}\,\mathbf{n} \,.
\end{equation}
Both $\dot{\theta}$ and $\mathbf{n}$ can be computed from $R$ using the generic procedure through quaternions. Introducing the rotation quaternion
\begin{equation}
   q = a + bi + cj + dk
   \hspace{0.2cm}\text{where}\hspace{0.2cm}
   i^2=j^2=k^2=ijk=-1\,,
\end{equation}
we have
\begin{equation}
   a = \cos\frac{\theta}{2}\,,
   \hspace{0.3cm}
   b = n_x\sin\frac{\theta}{2}\,,
   \hspace{0.3cm}
   c = n_y\sin\frac{\theta}{2}\,,
   \hspace{0.3cm}
   d = n_z\sin\frac{\theta}{2}\,,
\end{equation}
leading to
\begin{equation}
   \mathbf{\Theta} = \frac{2\dot{a}}{a^2-1}
   \begin{pmatrix}
      b\\
      c\\
      d
   \end{pmatrix}
   \hspace{0.3cm}
   \text{for }a\neq 1\ (\text{i.e., } \theta\neq 0).
\end{equation}
Each component $(a,b,c,d)$ of $q$ has a simple expression in terms of the components of the matrix $R$. The derivative $\dot{R}$ of the matrix $R$, required to compute $\dot{a}$, is obtained using the chain rule.
   
\end{document}